\documentclass[aps,prb,floatfix,twocolumn,superscriptaddress,showpacs,amsmath,amssymb]{revtex4}
\usepackage{graphicx}
\usepackage{epstopdf}
\usepackage{epsfig} 
\usepackage[applemac]{inputenx}     
\usepackage{amsmath}
\usepackage{bbm}

\allowdisplaybreaks

\newcommand{\be}{\begin{equation}}
\newcommand{\ee}{\end{equation}}
\newcommand{\bea}{\begin{eqnarray}}
\newcommand{\eea}{\end{eqnarray}}
\newcommand{\ba}{\begin{eqnarray*}}
\newcommand{\ea}{\end{eqnarray*}}
\newcommand{\dagga}{{\phantom{\dagger}}}

\newcommand{\bq}{\mathbf{q}}

\newcommand{\bk}{\mathbf{k}}

\newcommand{\bp}{\mathbf{p}}

\newcommand{\br}{\mathbf{r}}

\newcommand{\dis}{\displaystyle}

\newcommand{\fract}[2]{\frac{\dis #1}{\dis #2}}

\newcommand{\eqn}[1]{(\ref{#1})}

\newcommand{\bw}{\begin{widetext}}
\newcommand{\ew}{\end{widetext}}
\newcommand{\esp}[1]{\text{e}^{#1}}
\newcommand{\mH}{\mathcal{H}}
\newcommand{\bnot}{\mathbf{0}}

\begin{document}

\title{Non-equilibrium and non-homogeneous phenomena around a first-order quantum phase transition}

\author{Lorenzo Del Re} 
\affiliation{International School for
  Advanced Studies (SISSA), and CNR-IOM Democritos, Via Bonomea
  265, I-34136 Trieste, Italy}
\author{Michele Fabrizio} 
\affiliation{International School for
  Advanced Studies (SISSA), and CNR-IOM Democritos, Via Bonomea
  265, I-34136 Trieste, Italy} 
\author{ Erio Tosatti}
\affiliation{International School for
  Advanced Studies (SISSA), and CNR-IOM Democritos, Via Bonomea
  265, I-34136 Trieste, Italy}
  \affiliation{International Center for Theoretical Physics (ICTP), Strada Costiera 11, I-34151 Trieste, Italy }

\date{\today} 


\begin{abstract}
We consider non-equilibrium phenomena in
a very simple model that displays a zero-temperature first-order phase transition.  
The quantum Ising model with a four-spin exchange is adopted as a general representative of 
first-order quantum phase transitions that belong to the Ising universality class, 
such as for instance the order-disorder ferroelectric transitions, and possibly first-order $T=0$ Mott transitions. 
In particular, we address quantum quenches in the exactly solvable limit of infinite connectivity and show that, within 
the coexistence region around the transition, the system can remain trapped in a metastable phase, as long as 
it is spatially homogeneous so that
nucleation can be ignored.
Motivated by the physics of nucleation, we then study in the same model static but inhomogeneous phenomena that take place at surfaces and interfaces.
The first order nature implies that both phases remain locally stable across the transition, and with that the possibility of a metastable wetting layer 
showing up at the surface of the stable phase, even at $T$=0.  We use mean-field theory plus quantum fluctuations in the 
harmonic approximation to study  quantum surface  wetting.  

\end{abstract}
\maketitle

\section{Introduction}
The last few years have witnessed a growing interest in  
the non-equilibrium dynamics of systems driven across quantum phase transitions\cite{Altman-review,Wolf-review,Polkovnikov2011}, largely 
but not exclusively
stimulated by experiments on ultracold atoms\cite{bloch_review}.  
A rich 
theoretical activity thus focused on such systems, for example on the superfluid to Mott insulator transition in the one-dimensional Bose-Hubbard model\cite{Greiner_nature_02,
kollath_07,bakr_2010,Carleo} or  on the ferromagnetic to paramagnetic transition in the quantum Ising chain\cite{Nagerl2013,rossini_09,Calabrese&Essler2011}. In addition, 
studies have dealt with
quantum phase transitions 
in solid state systems, 
experimentally accessed by time-resolved pump-probe spectroscopies. That is for instance the case of the metal-insulator Mott transition in the Fermi-Hubbard model\cite{Werner-PRL,Schiro'}. This is a
prototypical model of strongly-correlated materials, whose rich phase diagram offers an ideal playground for out-of-equilibrium 
continuous 
phase transitions \cite{CavalleriVO2-2001,KublerVO2-2007,TakuboPrCaMnO3-2008,
BeaudLaCaMnO3-2009,
Cavalleri1TTaS2-2011,CavalleriNdSrMnO3-2011,
CavalleriLaSrMnO-2011,LiuVO2-2012,
Brazovskii2014,WolfVO2-2014}. 

However, phase transitions in real materials, 
including quantum ones,
are far more commonly discontinuous, while most of theoretical analyses focused so far on continuous quantum phase transitions. 
It is therefore important to identify and describe what 
novel features are brought about by the first order character and how they might show up in experiments. 

Here we take a first step in this direction by studying a very simple infinitely-connected model 
that displays a zero-temperature first-order transition. 
The simplicity of the model allows for exact results even in out-of-equilibrium conditions and may serve both as a reference example as well as a starting point 
for the successive addition of 
quantum fluctuations, which we shall partly undertake 
when dealing with
interface wetting. 

\subsection{Outline and summary of the main results}

The paper is divided into two parts. In the first, which comprises sections~\ref{The model} and \ref{Quantum quench}, we 
study the out of equilibrium dynamics of an exactly solvable infinitely connected quantum Ising model with a quartic spin-spin 
exchange. This model, as function of the transverse field $h$, displays a zero-temperature discontinuous transition from an ordered phase at $h<h_c$, where $Z_2$ symmetry is spontaneously broken, to a disordered one at $h>h_c$. A first-order quantum Ising phase transition should be relevant for the order-disorder para- to ferro-electric transition, but we speculate it might be representative also of discontinuous Mott transitions, as discussed in section~\ref{Links to the Mott transition}.
Because of the first order character, around $h_c$ there is coexistence between the two different phases, one being stable and the other metastable. The non-equilibrium protocol that we implement consists in a sudden quench of the transverse field, $h=h_i\to h_f$, starting from the ordered phase, i.e. with the initial $h_i<h_c$. We show that, because of coexistence, the system can remain trapped into an ordered state even though $h_f>h_c$. In this very peculiar solvable limit such result arises from a nice property of the microcanonical ensemble accessible by the unitary time-evolution. Indeed, in a certain range of transverse fields around $h_c$, the many-body spectrum includes a region of (extensive) eigenvalues where symmetry variant and invariant eigenstates coexist.  
Therefore, even though the low energy eigenvalues are symmetry invariant, i.e. the ground state is disordered, if the initial wavefunction has a finite overlap with the symmetry variant eigenstates the unitary evolution does not restore the $Z_2$ symmetry. In section~\ref{A non-equilibrium pathway} we further exploit this property showing that an external stimulus that acts for a finite time can drive the stable phase into the metastable one. 

A finite-temperature first-order phase transition can still show criticality due to wetting at an interface, a phenomenon that has been extensively studied since more than thirty years~\cite{Cahn1977,Lipowsky1982}. Not much it is known about wetting when the first order transition occurs instead at zero temperature. One can invoke the quantum-classical mapping and reasonably guess that the critical properties at zero temperature in $d$ dimensions should correspond to those of classical wetting in $d+1$ dimensions, actually $d+z$ where $z$ is the dynamical critical exponent.~\cite{Hertz} However, we think it is still worth uncovering what exactly wetting means in a quantum model that undergoes a first order transition as function of a Hamiltonian parameter, which is the content of the second part of this paper, section~\ref{Interface phenomena}. We first study wetting in the simplest extension of the infinitely connected quantum Ising model that consists of a slab where the spins are infinitely connected within each layer, which is however only coupled to its nearest neighbours. This model thus effectively maps onto a one-dimensional spin system, where each spin represents the total spin of each individual layer, hence it is of the order of its number of sites.  A semiclassical analysis of this model, which corresponds to the conventional spin-wave approximation, as shown in the Appendix and justified by the fact that the spin magnitude is large, shows that the gapped spin-wave excitation spectrum develops a bound state localised just at the wetting interface. We find that this bound state is unique and its energy vanishes at the first order transition. Because of the infinite connectivity we find, not surprisingly, that the critical properties of the saddle point are mean-field like and correspond to classical wetting in high dimensions. However the singular behaviour of quantum fluctuations is different from that of classical fluctuations at finite temperature. \\
In section~\ref{Capillary waves at the interface} we briefly discuss by means of spin-wave theory what should change in a more realistic three dimensional system where each layer is not infinitely stiff because of a finite connectivity. In this case the bound state develops into a branch of excitations that are still localised around the position of the wetting interface but disperse in the perpendicular direction. This surface, or, better, capillary wave is coherent, being below the continuum of bulk excitations, and goes soft at the first order transition. However such softening does not produce the logarithmic singularities that in the classical case signal interface roughening because of the extra dimension brought by quantum mechanics. \\
The final section~\ref{Conclusions} is devoted to concluding remarks, especially concerning our speculated relationship between a first-order zero-temperature Ising transition and first-order Mott transitions.  

\section{The model}
\label{The model}

Spin models on completely connected graphs have been largely studied in the context of continuous quantum phase transitions\cite{Semerjian,SciollaBiroli_short,SciollaBiroli_long,Matteo-ramps,Giacomo-PRB},  providing results that are quite instructive.  Our goal now is to move on to first order quantum phase transitions.   

We consider the quantum Ising model 
of  spins 1/2
on an $N$-site lattice 
\bea\label{model}
\mH &=& - \sum_{i,j}\,J^{(2)}_{ij}\;\sigma^z_i\,\sigma^z_j 
- h\,\sum_i\,\sigma^x_i
 \nonumber\\
&&  \qquad - \sum_{i,j,k,l}\,J^{(4)}_{ijkl}\;\sigma^z_i\,\sigma^z_j\,
\sigma^z_k\,\sigma^z_l\nonumber\\
&=& \bigg(-\frac{1}{N}\,J^{(2)}\,\big(\sigma^z_\bnot\big)^2\, - h\,\sigma^x_\bnot - \frac{1}{N^3}\,J^{(4)}\,
\big(\sigma^z_\bnot\big)^4 \bigg)\nonumber\\
&&  +\bigg(-\frac{1}{N}\,\sum_{\bq\not = \bnot}\,J^{(2)}_\bq\; \sigma^z_\bq\,\sigma^z_{-\bq}\nonumber\\
&& -\frac{1}{N^3}\,\sum_{(\bk\bp\bq)\not=(\bnot\bnot\bnot)}\,J^{(4)}_{\bk\bp,\bq}\,
\sigma^z_{\bk}\,\sigma^z_{\bp+\bq}\,\sigma^z_{-\bp}\,\sigma^z_{-\bk-\bq}\bigg)\nonumber\\
&=& \mH_\bnot + \delta\mH,\label{Ham} 
\eea      
where $\sigma^a_i$, $a=x,y,z$, are Pauli matrices, $J^{(2)}= J^{(2)}_\bnot$ and $J^{(4)}=J^{(4)}_{\bnot\bnot,\bnot}$ are the Fourier transforms of the two-spin and four-spin exchange constants at zero momentum. 
The multi-spin exchange is known to lead to first order transitions. For instance, the Ising model with four spin interaction has been invoked to describe the first order paraelectric-ferroelectric phase transition, especially in order-disorder ferroelectrics.~\cite{Ferroelectrics-1}
In Eq.~\eqn{Ham}, $\mH_\bnot$ includes only Fourier components at $\bq=\bnot$, while $\delta\mH$ embodies all  remaining terms. In the limit of long-range exchange we can 
discard $\delta \mH$ and just study the mean-field Hamiltonian $\mH_\bnot$. Successively, we could treat $\delta \mH$ perturbatively, for instance in the spin-wave approximation\cite{Matteo-ramps}, which would amount to add spatial quantum fluctuations to the mean-field results.

The Hamiltonian $\mH_\bnot$, which corresponds to an infinitely connected Lipkin-Ising model, was studied in non-equilibrium conditions by Bapst and Semerjian in Ref.~\onlinecite{Semerjian}, see also 
Ref.~\onlinecite{SciollaBiroli_long}. 
Here we shall closely follow their work and extend it to account for 
a larger variety of out-of-equilibrium situations. 
 
\subsection{Equilibrium properties of $\mH_\bnot$} 
We start from the equilibrium phase diagram of $\mH_\bnot$. 
We set $J^{(4)}$ as our energy unit, and define 
$\gamma=J^{(2)}/J^{(4)}$ and $h/J^{(4)}\to h$.    

Observing that $\mH_\bnot$ commutes with the total spin
\be
\mathbf{S}^2 = \bigg(\frac{1}{2}\,\sum_i\,\boldsymbol{\sigma}_i\bigg)^2 = S(S+1),
\ee
the Hilbert space decomposes into subspaces with fixed 
total spin $S=0,1,\dots,M$, assuming $N=2M$ an even integer. Each subspace 
comprises eigenstates of the Hamiltonian
\be
\mH_\bnot(S) = -\frac{4}{N}\,\gamma\,\Big(S^z\Big)^2 
-\frac{16}{N^3}\,\Big(S^z\Big)^4 - 2h\,S^x,\label{H0}
\ee
each with degeneracy  
\be
g(S) = \binom{2M}{M+S} - \binom{2M}{M+S+1},
\ee
which is the number of ways to couple $N$ 1/2 spins into total spin $S$. If we define 
$S = M\,s$, then, in the thermodynamic limit $M\to\infty$, $s$  effectively becomes a continuous variable, $s\in[0,1]$, and the partition function can be evaluated semiclassically\cite{Lieb} 
\bea
Z &=& \sum_S\,g(S)\,\sum_{S^z=-S}^S\, \langle S,S^z\mid 
\esp{-\beta\,\mH_\bnot(S)}\mid S,S^z\rangle\label{Z}\\
&\simeq& \int_0^{1}\!\!\!ds \, \fract{M\big(2M s+1\big)}{4\pi}\int \!d\!\cos\theta\,d\phi \;
\esp{-N\beta\,F(\theta,\phi,s)},\nonumber
\eea     
where the semiclassical free-energy density reads
\bea
F(\theta,\phi,s) &=& -\gamma\,s^2\!\cos^2\theta 
-s^4\!\cos^4\theta\nonumber\\
&& \!\!- h\,s\sin\theta\,\cos\phi 
- T\,\mathcal{S}(s),\label{F}
\eea  
with entropy 
\bea
\mathcal{S}(s) &=& \frac{1}{2M}\,\ln g\big(Ms)\big) \simeq - \fract{1+s}{2}\,\ln\fract{1+s}{2}\nonumber\\
&& - \fract{1-s}{2}\,\ln\fract{1-s}{2} + O\big(M^{-1}\big).
\label{entropy}
\eea
For $M\to\infty$ we can approximate the integral by its saddle point value, which corresponds to $\cos\phi=1$ and, upon defining  
$m=\cos\theta\in[-1,1]$, to the minimum of 
\bea
F(m,s) &=& -\gamma\,s^2 m^2 
-s^4 m^4- h s\,\sqrt{1-m^2} -T\,\mathcal{S}(s)
\nonumber\\
&=& E(m,s) -T\,\mathcal{S}(s),\label{F(m)}
\eea  
with respect to both $s$ and $m$. 

At fixed $s$, the extrema of $E(m,s)$ satisfy
\be
m\bigg[h-\sqrt{1-m^2}\;\Big(2\gamma s + 4s^3 m^2\Big)\bigg]=0,
\ee
so that $m=0$ is always an extremum, actually a maximum whenever 
$4\gamma^2 s^2 > h^2$. We define $t\equiv \sqrt{1-m^2} \in [0,1]$ so that, apart from $m=0$, the other extrema satisfy the cubic equation
\bea
t^3 -\Big(1+\frac{\gamma}{2s^2}\Big)\,t
+ \fract{h}{4s^3} \equiv t^3 -a\,t
+ b = 0,\label{uno}
\eea
whose roots are acceptable solutions only if real, positive and smaller than one. Solving Eq.~\eqn{uno}, we find three distinct regimes:
\begin{itemize}
\item[{\bf 1.}] If 
\be
h^2 \geq \frac{8}{27}\,\big(2s^2+\gamma\big)^3,\label{1a}
\ee
or 
\be
\Big(4\gamma^2 s^2\leq h^2 \leq \frac{8}{27}\,\big(2s^2+\gamma\big)^3\Big)\,\land\,\Big(\gamma\geq 4s^2\Big),\label{1b}
\ee
there is only one minimum at $m=0$.
\item[{\bf 2.}] If 
\be
h^2 \leq 4\gamma^2 s^2,\label{2}
\ee
then $m=0$ is a maximum and there are two equivalent minima at
\be 
m_\pm =\pm m_* =\pm \sqrt{1-t_*^2},\label{m_pm}
\ee
where 
\be
t_* = 2\sqrt{\fract{a}{3}}\;\cos\zeta,\label{t*}
\ee
with 
\be
\zeta = \fract{2\pi}{3}-
 \frac{1}{3}\,\mbox{tan}^{-1} \fract{\sqrt{\fract{4 a^3}{27}-b^2}}{b}
\in \Big[\frac{\pi}{3},\frac{\pi}{2}
\Big].\label{zeta}
\ee
Because of the finite magnetisation $m_*$, the $Z_2$ symmetry is spontaneously broken. 
\item[{\bf 3.}] Finally, when 
\be
\Big(4\gamma^2 s^2\leq h^2 \leq \frac{8}{27}\,\big(2s^2+\gamma\big)^3\Big)\,\land\,\Big(\gamma\leq 4s^2\Big),\label{3}
\ee
there are three minima, one at $m=0$ and the two equivalent 
symmetry-breaking minima 
Eq.~\eqn{m_pm}, with the same $t_*$ as in Eq.~\eqn{t*}. The absolute minimum depends on the value of $E(0,s)$ with respect to 
$E\big(\pm m_*,s\big)$.
\end{itemize}
It follows that each individual subspace at fixed $s$ undergoes a quantum phase transition from an ordered ground state with finite magnetisation below a critical $h_*(s)$ to a disordered one above. The critical $h_*(s)$ is lower the lower $s$; the transition is first order 
if $\gamma\leq 4s^2$ and second order otherwise. Conversely, at fixed 
$h$, there exists a critical $s_*(h)$ such that the ground states of all subspaces with 
$s\geq s_*(h)$ are ordered while those of the subspaces with $s<s_*(h)$  are disordered.  

\begin{figure}[t]
\centerline{\includegraphics[width=9cm]{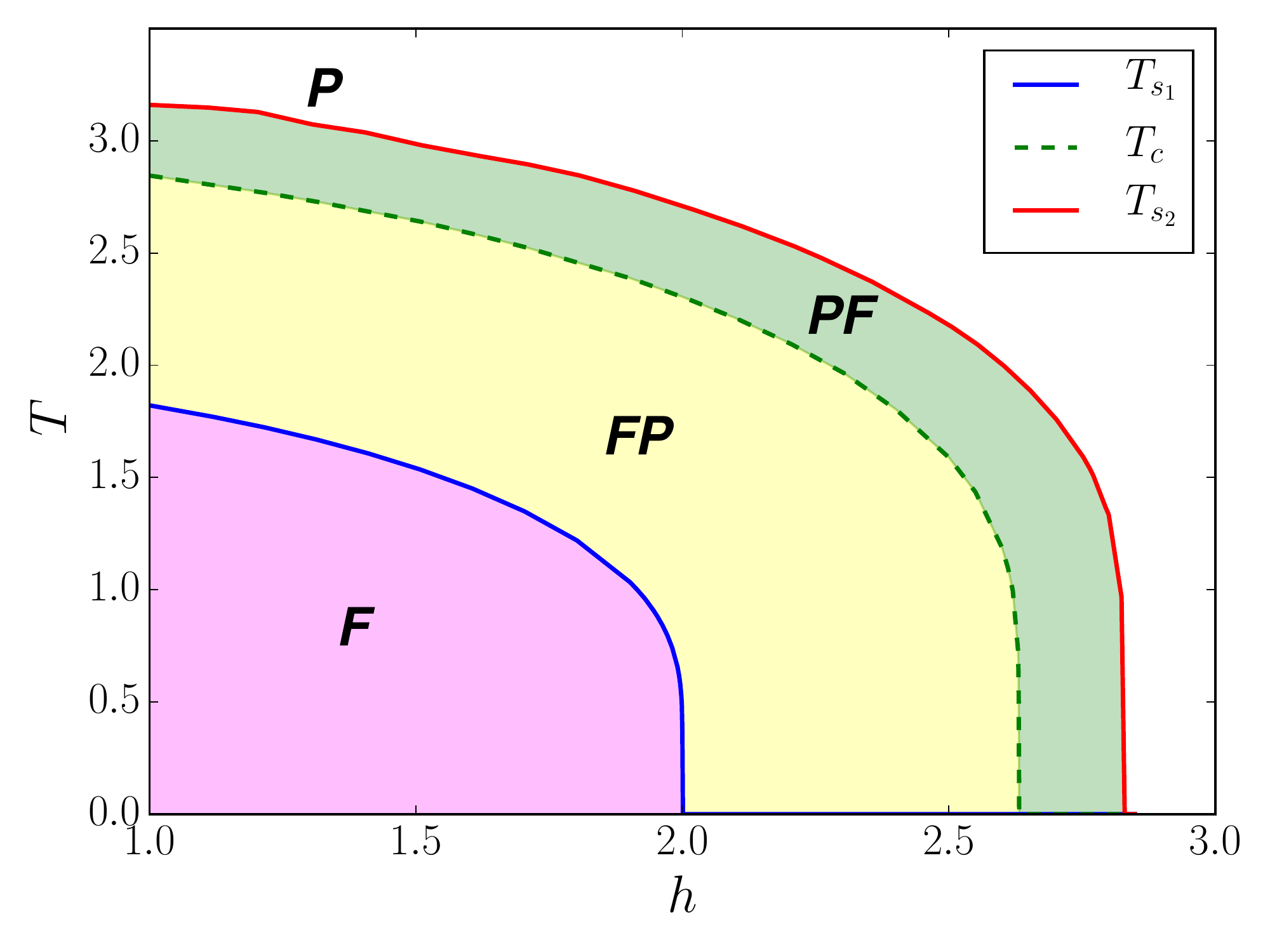}}
\caption{(Color online) Phase diagram of the model $\mH_0$, Eq.~\eqn{Ham}, at $\gamma=1$. In the region labelled as {\bf F} there is only a ferromagnetic free-energy minimum, i.e. the $Z_2$ symmetry is broken. Conversely, in the region {\bf P} there is only a paramagnetic minimum. The coexistence regions are {\bf FP} and {\bf PF}, 
where in {\bf FP} the absolute minimum is ferromagnetic, while 
in {\bf PF} is paramagnetic. The F-P phase transition occurs where the free energies cross along the dotted line $T_{c}$ separating {\bf FP} from {\bf PF}.
The solid lines $T_{s_1}$ between  {\bf F} and {\bf FP}, and $T_{s_2}$ between {\bf P} and {\bf PF},  are spinodal lines where an additional metastable free energy minimum,  P and F, respectively, appears besides the stable one, F and P respectively.     } 
\label{phase-diagram}
\end{figure}

Let us analyse the above results more in detail. At $T=0$, the 
ground state is in the $s=1$ subspace. We shall always assume that 
\be
\gamma < 4, \label{assumption}
\ee
so that, with $s=1$, the only possibilities are Eqs.~\eqn{1a}, 
\eqn{2} and \eqn{3}. It thus follows that at $h^2\leq 4\gamma^2$, 
the symmetry is broken and $m=\pm m_*$. In the region 
\be
4\gamma^2 \leq h^2 \leq \frac{8}{27}\,\big(2+\gamma\big)^3,
\ee
there is coexistence between a symmetry broken solution, $m=\pm m_*$, and a symmetry invariant one, $m=0$. The two solutions cross at a 
critical $h_c=h_*(s=1)$, which signals a first order quantum phase transition. Finally, for 
\be
h^2 \geq \frac{8}{27}\,\big(2+\gamma\big)^3,
\ee
the symmetry broken solution is no longer stable. 

Suppose now that $h<h_c$, so that at $T=0$ the $Z_2$ symmetry is spontaneously broken. At finite temperature, $T\not= 0$, because of the entropy term Eq.~\eqn{entropy}, 
the free energy minimum is obtained, see Eq.~\eqn{F(m)}, for a value of the total spin $s=s(T)<1$, which diminishes with increasing $T$. Therefore, there exists a critical temperature $T_c$ at which 
$s\big(T_c\big) \leq s_*(h)$, 
where the symmetry is recovered. Specifically, if we assume Eq.~\eqn{assumption}, the transition remains first order at any finite $T$. The phase diagram at $\gamma=1$ is shown in Fig.~\ref{phase-diagram} for $h\geq 1$.

\begin{figure}[t]
\centerline{\includegraphics[width=9cm]{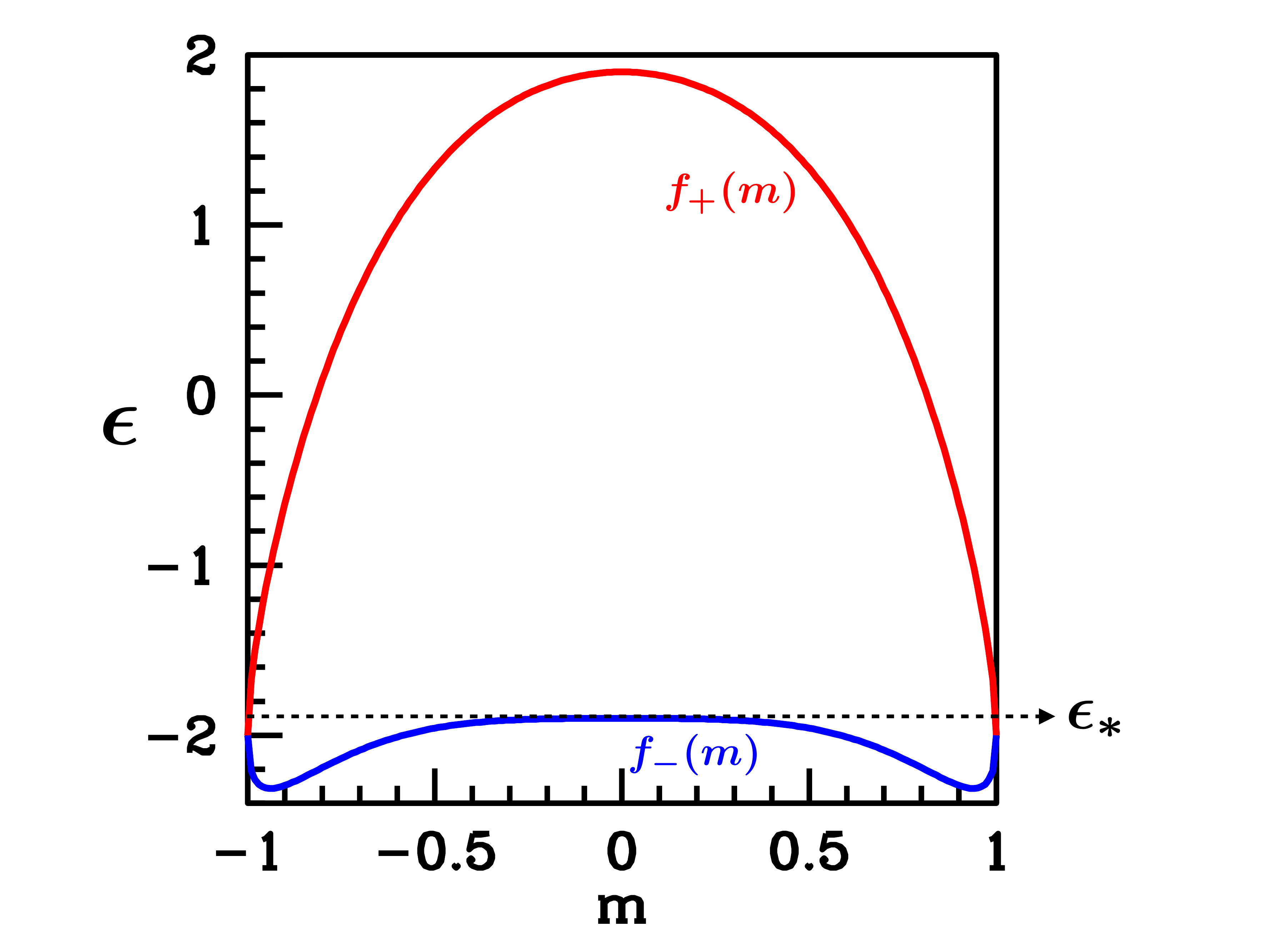}}
\caption{(Color online) The function $f_+(m)$ and $f_-(m)$ for $s=1$, $\gamma=1$ and $h=1.9$, which correspond to the case in which the ground state breaks 
spontaneously $Z_2$. The regions above $f_+(m)$ and below $f_-(m)$ are unaccessible when $N\to\infty$. Also shown is the energy threshold 
$\epsilon_*$ above which symmetry is restored.} 
\label{h1.9}
\end{figure}

\subsection{Links to the Mott transition}
\label{Links to the Mott transition}
As was said above, the Ising model with a four spin exchange, Eq.~\eqn{Ham}, describes the first order paraelectric-ferroelectric phase transition, especially in order-disorder ferroelectrics.~\cite{Ferroelectrics-1}
Here we 
propose a link to another equally interesting class of first order phase transitions: the Mott metal-to-insulator transition. 
In 1979, Castellani and coworkers~\cite{Castellani} suggested that the finite temperature transition from a paramagnetic metal to a paramagnetic Mott insulator in V$_2$O$_3$ 
should belong to the Ising universality class. This conjecture was later put by Kotliar, Lange and Rozenberg~\cite{Gabi2000} on firmer theoretical grounds after the development of dynamical mean field theory (DMFT).~\cite{DMFT} It was finally experimentally 
confirmed both in V$_2$O$_3$~\cite{Limelette2003} and in organic compounds~\cite{Organic2004}. 

More recently, the half-filled single-band Hubbard model on an infinitely coordinated lattice -- the limit where DMFT is exact -- was rigorously proven to be mappable onto a model of Ising spins coupled to non-interacting electrons.~\cite{MarcoPRB} In this mapping, 
the metal phase translates into an ordered phase with broken $Z_2$ symmetry and the zero-temperature second-order Mott transition corresponds to recovering that symmetry. The mean-field decoupling of the Ising degree of freedom from the electronic ones,
which provides a qualitatively correct description of the model~\cite{Baruselli,Rok} 
(and is also equivalent to the Gutzwiller variational wavefunction~\cite{MarcoPRB}) 
leads to an effective mean-field Hamiltonian for the Ising spins that is exactly $\mH_\bnot$ in Eq.~\eqn{Ham} 
with $J^{(4)}=0$, i.e. a standard infinitely-connected Ising model in a transverse field. 

If we assume the above Ising dictionary to hold even when the zero-temperature Mott transition becomes 
first order, then we can conjecture that the  Hamiltonian $\mH_0$ with $J^{(4)}\not =0$ might provide a sensible 
mean-field description of that transition too. Therefore, even though we shall formally deal with the spin model 
Eq.~\eqn{Ham}, we will often refer to the above mapping and regard the ferromagnetic phase as the metal, 
and the paramagnetic one as the Mott insulator.  

\section{Time inhomogeneity: quantum quench}
\label{Quantum quench}
We  start  by highlighting a peculiarity of the infinitely connected 
Hamiltonian $\mH_\bnot$ Eq.~\eqn{Ham}, see also Eq.~\eqn{H0}. 
Since the total spin $S$ is a conserved quantity, one can either consider the "grand canonical" ensemble of Eq.~\eqn{Z}, or focus on a "canonical ensemble" at fixed $S$. Each individual subspace at fixed $S=Ms$ 
undergoes a quantum phase transition by increasing $h$ above $h_*(s)$, whereas, if $h\leq h_*(s)$, it has no transition upon rising $T$. Only in the "grand canonical" ensemble 
a finite temperature phase transition is possible because the  
entropy term Eq.~\eqn{entropy} causes subspaces with progressively lower total spin $s$ 
to dominate as the temperature rises.
\begin{figure}[bht]
\centerline{\includegraphics[width=9cm]{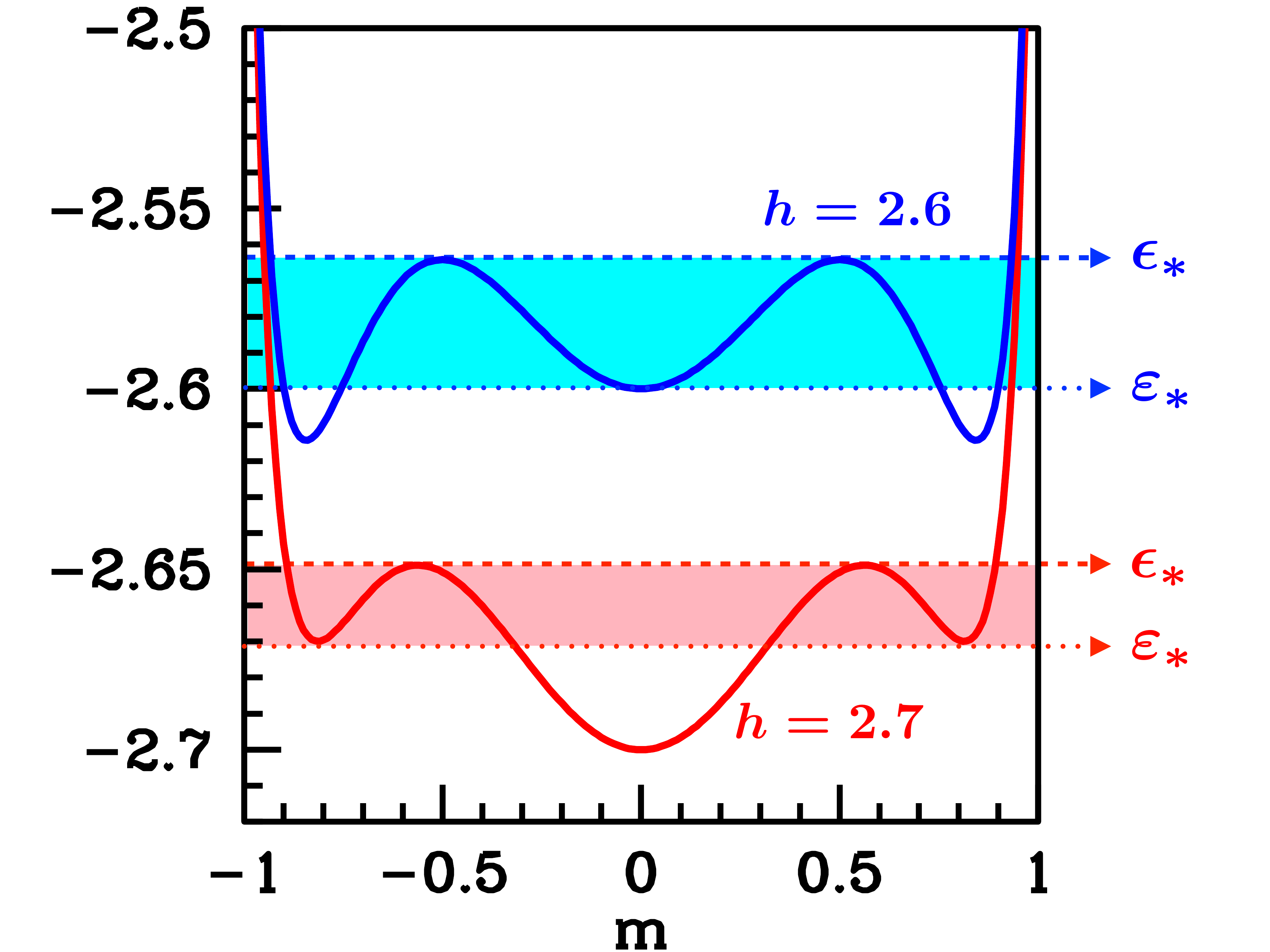}}
\caption{(Color online) The function $f_-(m)$ for $s=1$, $\gamma=1$ and two values of $h$: 
$h=2.6$ (blue curve on top) and $h=2.7$ (red curve at bottom), which correspond to values just across the 
first order phase transition, $h_c\simeq 2.63113$. Even in these cases there is a threshold energy $\epsilon_*$ above which all eigenstates are symmetry invariant. In addition, there is coexistence of symmetry breaking and symmetry invariant eigenstates in the interval $[\varepsilon_*,\epsilon_*]$.  } 
\label{mix}
\end{figure}

A phase transition within a given subspace may  on the contrary still occur in the microcanonical ensemble, i.e. at fixed total energy. Without repeating the calculations of Ref.~\onlinecite{Semerjian}, we just recall that the allowed eigenvalues $E=N\epsilon$ of $\mH_\bnot$, Eq.~\eqn{H0}, in the semiclassical $N\to\infty$ limit are restricted 
within the interval 
\be
\min_m\Big(f_-(m,s)\Big)\leq \epsilon\leq \text{Max}_m\Big(f_+(m,s)\Big),
\ee
where 
\ba
f_\pm(m,s) &=& -\gamma s^2\,m^2 - s^4\,m^4 \pm h\,s\,\sqrt{1-m^2}\, , .
\ea
Conversely, eigenstates with eigenvalues $E=N\epsilon$ and total spin $S$ will have components on states $\mid S^z,S\rangle$, where $S= s \,N/2$ and $S^z= S\,m$, only for values of $m$ such that 
$f_-(m,s) \leq \epsilon$ and $f_+(m,s) \geq \epsilon$. Outside that interval, the eigenfunctions decay exponentially fast in $N$. 

In Fig.~\ref{h1.9} we show $f_+(m,s)$ and $f_-(m,s)$ for $s=1$, $\gamma=1$ and $h=1.9$, 
which corresponds to the case in Eq.~\eqn{2} with an ordered ground state. 
As discussed in Refs.~\onlinecite{Semerjian}, all eigenstates with 
eigenvalues $\epsilon\leq \epsilon_*$, (see figure)  are doubly degenerate and symmetry non-invariant, 
while those with $\epsilon > \epsilon_*$ are non-degenerate and symmetry invariant. There is therefore 
a symmetry breaking edge\cite{Giacomo-PRB} defined as the critical energy $E_*=N\,\epsilon_*$ above 
which symmetry is restored in the microcanonical ensemble. More interesting is the behavior in the coexistence region of Eq.~\eqn{3}. In Fig. \ref{mix} we show 
$f_-(m,s)$ still for $s=1$ and $\gamma=1$, 
and for two different values of $h=2.6$ and $h=2.7$ 
on both sides of 
the first order phase transition. In this case 
there are two relevant energy thresholds. For instance, when $h=2.6$, on the ferromagnetic side, for $\epsilon\leq \varepsilon_*$ 
all eigenstates are symmetry breaking. In the region $\varepsilon_*\leq \epsilon\leq \epsilon_*$ there are 
both symmetry non-invariant and invariant eigenstates. Finally, when $\epsilon>\epsilon_*$, only symmetry 
invariant eigenstates exist. On the contrary, on the paramagnetic side of the transition at $h=2.7$, 
the low energy eigenstates, $\epsilon\leq \varepsilon_*$, are symmetry invariant, while for 
$\varepsilon_*\leq \epsilon\leq \epsilon_*$ there is coexistence.

The semiclassical equation of motion for the magnetization $m$ is equivalent to the Hamilton-Jacobi 
equation corresponding to the 
classical Hamiltonian\cite{SciollaBiroli_long,Giacomo-PRB} 
\be
\mathcal{H}(m,\phi) = -\gamma\,s^2\,m^2 - s^4\,m^4 - h\,s\,\sqrt{1-m^2}\;\cos 2\phi,\label{classical-H}
\ee
with $m$ and $\phi$ conjugate variables.  
Since the energy is conserved, if initially the state has energy 
$\epsilon$ then
\be
\dot{m} = \mp 2\sqrt{\Big(f_+(m,s)-\epsilon\Big)\Big(\epsilon-f_-(m,s)\Big)}\;,\label{mdot}
\ee
which explicitly shows that, if $\epsilon\geq \epsilon_*$, even if initially $m(t=0) \not = 0$, during the evolution $m(t)$ oscillates between $\pm m_0$ where $f_-(\pm m_0,s)=\epsilon$, so that its time average 
\[
\overline{m}(\tau)= \fract{1}{\tau} \int_0^\tau\, dt\, m(t),
\]
vanishes over a period $\tau$. 
\begin{figure}[tbh]
\centerline{\includegraphics[width=9cm]{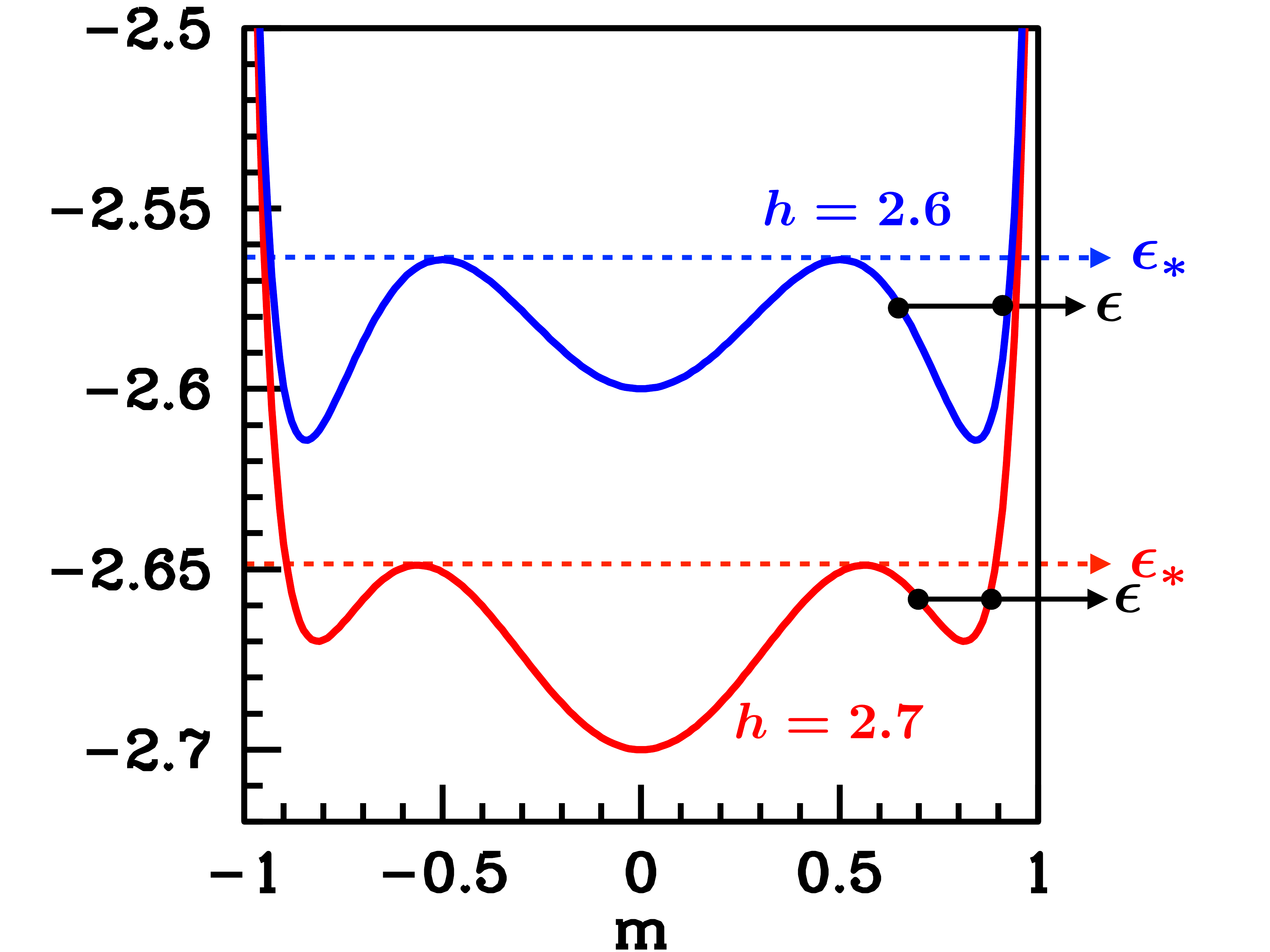}}
\caption{(Color online) As in Fig. \ref{mix}, showing two initial conditions identified by the energy 
$\epsilon$. In both cases the magnetization will oscillate between the two black points shown in the figure.  } 
\label{mix2}
\end{figure}

On the contrary, if initially $m(0)$ 
falls within one of the $f$-valleys at
finite magnetisation, see Fig.~\ref{mix2}, 
its time average remains finite until $\epsilon\leq \epsilon_*$, even if the actual minimum is at $m=0$, 
such as for 
$h=2.7$. As soon as $\epsilon$ exceeds $\epsilon_*$, 
the time average suddenly jumps to zero, signalling a first order dynamical restoration
of symmetry. 

\subsection{A non-equilibrium pathway}
\label{A non-equilibrium pathway}
Let us consider the Hamiltonian $\mathcal{H}(m,\phi)$ at $\gamma=1$ in the coexistence region for 
$h=2.631131\gtrsim h_c$,  
where 
the ground state is disordered, 
i.e. $m=0$, but there exists a metastable state at $m\simeq 0.836$, and add a perturbation
\be
\delta\mathcal{H}(\phi,t) = -\mu\,\theta\big(\tau-t\big)\,\phi,\label{perturbation}
\ee
which displaces the magnetisation and acts for a finite time $\tau$. The semiclassical equations of motion 
now read
\bea
\dot{m}(t) &=& -\fract{\partial \mathcal{H}(m,\phi)}{\partial \phi} + \mu\,\theta\big(\tau-t\big),
\label{t-m}\\
\dot{\phi}(t) &=& \fract{\partial \mathcal{H}(m,\phi)}{\partial m} - \eta\,\phi(t),\label{t-phi}
\eea
where $\eta$ is the strength of a dissipative process that we need to add 
in order
to force relaxation to a steady state.  
\begin{figure}[tbh]
\centerline{\includegraphics[width=0.95\columnwidth]{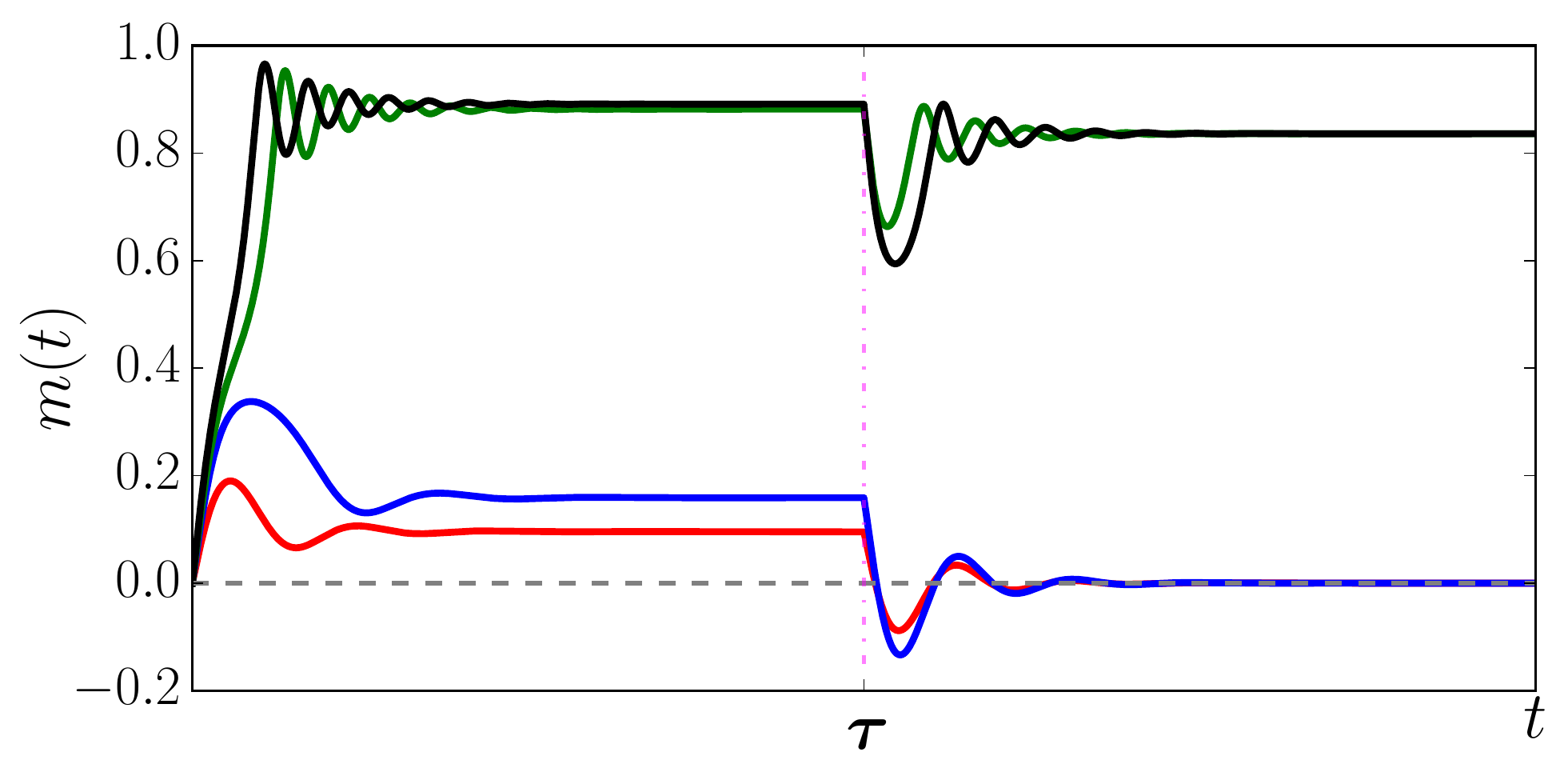}}
\caption{(Color online) Dynamical evolution given by Eq.~\eqn{t-m} and \eqn{t-phi} for increasing 
$\mu=0.4,~0.6,~0.7$ and 0.8 from bottom to top, with the dissipation parameter $\eta=1.5$. } 
\label{m-neq}
\end{figure}
In Fig.~\ref{m-neq} we show the non-equilibrium time-dependent magnetisation as derived by solving  
Eqs.~\eqn{t-m} and \eqn{t-phi}. 
At the initial time the system is in the equilibrium configuration, 
$m(0)=\phi(0)=0$, and is let evolve in the presence of the perturbation Eq.~\eqn{perturbation} for increasing 
$\mu$, curves from bottom to top. We observe that 
whereas for small $\mu$ the system returns to the ground state as expected after the switching off of the $\mu$ term, 
above a threshold value of $\mu$ it remains trapped into the metastable state 
corresponding to a side $f$-valley, 
a trapping all the more suggestive given such a simple and  dissipative model. 
This trapping is intimately connected with spatial uniformity, as we shall see.

\section{Space Inhomogeneity: Interface Phenomena}
\label{Interface phenomena}
The dynamical evolution depicted in Fig.~\ref{mix2} for $h=2.7$, or that in Fig.~\ref{m-neq}, where the system 
ends up trapped in a metastable symmetry breaking phase is appealing but not physical. In real systems, bubbles of the 
stable disordered phase nucleate by quantum fluctuations, and eventually grow at the expense of the metastable long-range-ordered phase, 
finally destroying it . 
Quantum nucleation has been extensively studied in a wide variety of contexts, ranging from cosmology to quantum liquids, and the literature on the argument is so vast that it is practically impossible to track it after the seminal works by Coleman~\cite{Coleman} and by Voloshin, Kobzarev and Okun'~\cite{Voloshin}. 
In the case of macroscopic magnets, which is more pertinent to the present work, we may refer to two review articles, Refs.~\onlinecite{Stamp1992} and \onlinecite{Owerre2015}, and references therein.

Our aim here will not be to explore quantum nucleation in the toy model Eq.~\eqn{Ham}, whose overall phenomenology is 
not going to differ much from the models studied for instance in Ref.~\onlinecite{Stamp1992}. However, nucleation implies a quantum interface between the stable and the metastable phase. The physics of this
type of interface, even when static, demands a separate study. Its results  could be of additional interest in view of our 
conjectured connection with electronic models displaying a zero-temperature first-order Mott transition.  

We already mentioned that the effective theory which describes within the Gutzwiller approximation the single-band Mott transition is  the Ising model Eq.~\eqn{Ham} with $J^{(4)}=0$.  In this particular context, Borghi {\sl et al.} studied~\cite{Borghi-PRL,Borghi-PRB} several quantum interfaces of direct physical interest, for instance between a Mott insulator and a metal, or between a correlated metal and the vacuum. In the latter case it was found that a poorly metallic, nearly insulating {\sl dead layer} appears at the metal surface, with a thickness controlled by the Mott transition correlation length. Because in that model the Mott transition is continuous, the correlation length can, close to the transition, be much larger than characteristic electronic length scales of the metal,  i.e. the Fermi wavelength or the screening length. An anomalously thick superficial dead layer devoid of measurable quasiparticle weight was indeed observed by surface sensitive photoemission  in V$_2$O$_3$~\cite{Marino2009}. However the Mott transition in V$_2$O$_3$ is first order, hence it is not guaranteed that the 
results of Borghi {\sl et al.}~\cite{Borghi-PRL,Borghi-PRB}, valid for a continuous transition, are representative. 

\begin{figure}[htb!]
\includegraphics[width = 0.99\columnwidth]{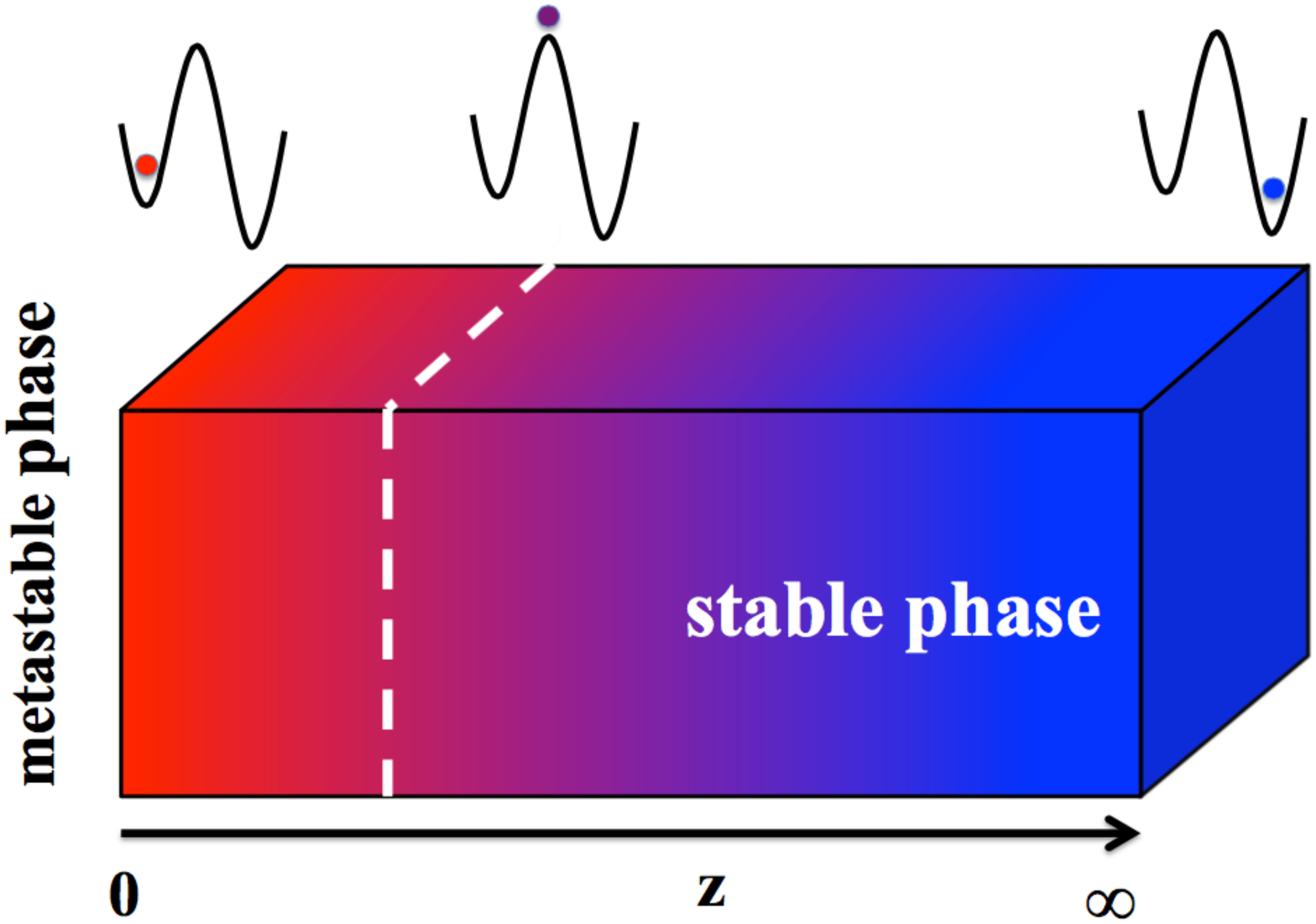}
\caption{\label{slab} The surface phenomena that we investigate. A semi-infinite slab $z\in [0,\infty]$ is constrained to lie at the surface $z=0$ in the metastable minimum, while, deep in the bulk, $z\gg 1$, it recovers its equilibrium phase. The metastable phase wets the surface layers. At a particular $z=z_*$ the system finds itself on the top of the potential barrier, as sketched in figure, which we shall associate with the position of the wetting interface.}
\end{figure}

Here we shall repeat the analyses of Refs.~\onlinecite{Borghi-PRL} and \onlinecite{Borghi-PRB} with 
$J^{(4)}\not =0$ and highlight the novel features brought by the first order character of the  transition. 
Specifically we shall study the situation, schematically drawn in Fig.~\ref{slab}, of a semi-infinite slab, 
$z\in [0,\infty]$, constrained to lie at the surface, $z = 0$, in the metastable minimum with $m(z=0)=m_0$. 
Deep in the bulk the slab flows back to its equilibrium phase, i.e. the absolute potential minimum at 
$m = m_\infty$, so that $m(z\to\infty) = m_\infty$. As discussed 
in section \ref{Links to the Mott transition} the specific case of $m_0=0$ and $m_\infty$ finite should translate precisely in an interface between the vacuum ($m_0=0$) and a 
correlated metal ($m_\infty\not = 0$).    

As expected close to any first order phase transition, such as melting,~\cite{Tartaglino2005} we will find that the metastable phase wets the surface layers.
That is, by moving from the "disordered" surface at $z$ = 0 into the "ordered" bulk, the bulk order parameter is not recovered in a simple exponential manner as in the second order case 
studied by Refs.~\onlinecite{Borghi-PRL}. Instead, a disordered 
film of small but finite thickness will form at the surface, inside which $m(z) \sim m_0 $, subsequently  turning from  $m_0 $
to  $m_\infty$ at the "wetting interface".  We shall define the position of the wetting interface with $z=z_*$ such that $m(z_*)$ corresponds to the top of the barrier between the two minima, as sketched in Fig.~\ref{slab}. In what follows, we will start with the classical equilibrium configuration and then study quantum fluctuations of this wetting interface. 

\subsection{The continuous limit of the slab}
In order to study interface phenomena we 
must
abandon the simple model $\mH_0$, which cannot describe spatial inhomogeneities, and consider the whole Hamiltonian $\mH_0+\delta\mH$ in Eq.~\eqn{Ham}. 
Alternatively, we may adopt a less rigorous approach and consider the model in a 
slab 
geometry. Each layer $i=1,\dots,L$ in the slab, with number of sites $A\to\infty$, is described by the infinitely connected Hamiltonian $\mH_0$ in the $S=A/2$ sector, and it is coupled to its adjacent layers $i-1$ and $i+1$ by a quadratic and a quartic term.
 
Specifically, the Hamiltonian per unit area reads  
\bea\label{lattice:model}
\mH &=& \sum_{i=1}^L \mathcal{H}_{0i} 
-\sum_i\,\bigg[ \gamma_\perp\,m_i\,m_{i+1}
+ \tilde{\gamma}_{\perp}m_i^2\,m_{i+1}^2\bigg]
,\;\;\;\;\;
\eea
where  
\begin{equation}
{\mathcal{H}}_{0i} =   -\gamma\,m_i^2 
-\tilde{\gamma}\,m_i^4 - h\,\sqrt{1-m_i^2}\;\cos 2\phi_i,
\end{equation}
with $-1\leq m_i\leq 1$ and $\phi_i$ conjugate variables, $\gamma_\perp$ and $\tilde{\gamma}_\perp$ the strengths of the quadratic and quartic interactions between adjacent  layers, respectively. 
The model thus describes a spin chain where the spin at site $i$ is actually the total spin of the individual layer $i$, hence its magnitude $S=A/2$ is large. Since each layer is infinitely connected, it is also infinitely stiff so that the model by construction cannot describe fluctuations perpendicular to the slab axis. 

Taking the continuum limit of Eq.~(\ref{lattice:model}) yields the following expression of the Hamiltonian:
\begin{eqnarray}\label{continuum:model}
\mathcal{H} &=& \int_0^\infty dz \bigg[- (\tilde{\gamma} + \tilde{\gamma}_\perp)m^4(z) - (\gamma+\gamma_\perp) \,m^2(z) +\nonumber \\ \nonumber \\&-& h\sqrt{1-m^2(z)}\cos2\phi(z) +  \nonumber \\ \nonumber \\
& -& \frac{\gamma_\perp}{2}m(z)\,\frac{\partial^2 m(z)}{\partial z^2} - \frac{\tilde{\gamma}_\perp}{2}m(z)^2\,\frac{\partial^2m(z)^2}{\partial z^2} \bigg] \\
&=& C + \int_0^\infty \!\!dz \bigg[ \fract{K_\perp\big(m(z)\big)}{2}\,\left(\fract{\partial m(z)}{\partial z}\right)^2 
+ V\big(m(z)\big)\nonumber\\
&&  \phantom{C+\int_0^\infty dz\;} + 2h\,\sqrt{1-m^2(z)}\;\sin^2 \phi(z) \bigg],\nonumber
\end{eqnarray}
where, as discussed earlier, we assumed that $m(0)=m_0$ and $m(\infty)=m_\infty$ are fixed , 
\[ 
C = -\frac{1}{4}\int_0^\infty dz\, \bigg[\gamma_\perp\,  \frac{\partial^2 m(z)^2}{\partial z^2}
+ \tilde{\gamma}_\perp\,  \frac{\partial^2 m(z)^4}{\partial z^2}\bigg],
\]
is a constant that depends on the boundary values and which we shall drop hereafter. We defined 
\bea
V(m) &=& -(\tilde{\gamma}+\tilde{\gamma}_\perp)\,m^4 - (\gamma + \gamma_\perp)\,m^2 
\nonumber\\
&& - h \sqrt{1-m^2},\label{V(m)}\\
K_\perp(m) &=& \gamma_\perp + 4\,\tilde{\gamma}_\perp\,m^2.\label{K(m)}
\eea
\begin{figure}[thb!]
\includegraphics[width = 0.99\columnwidth]{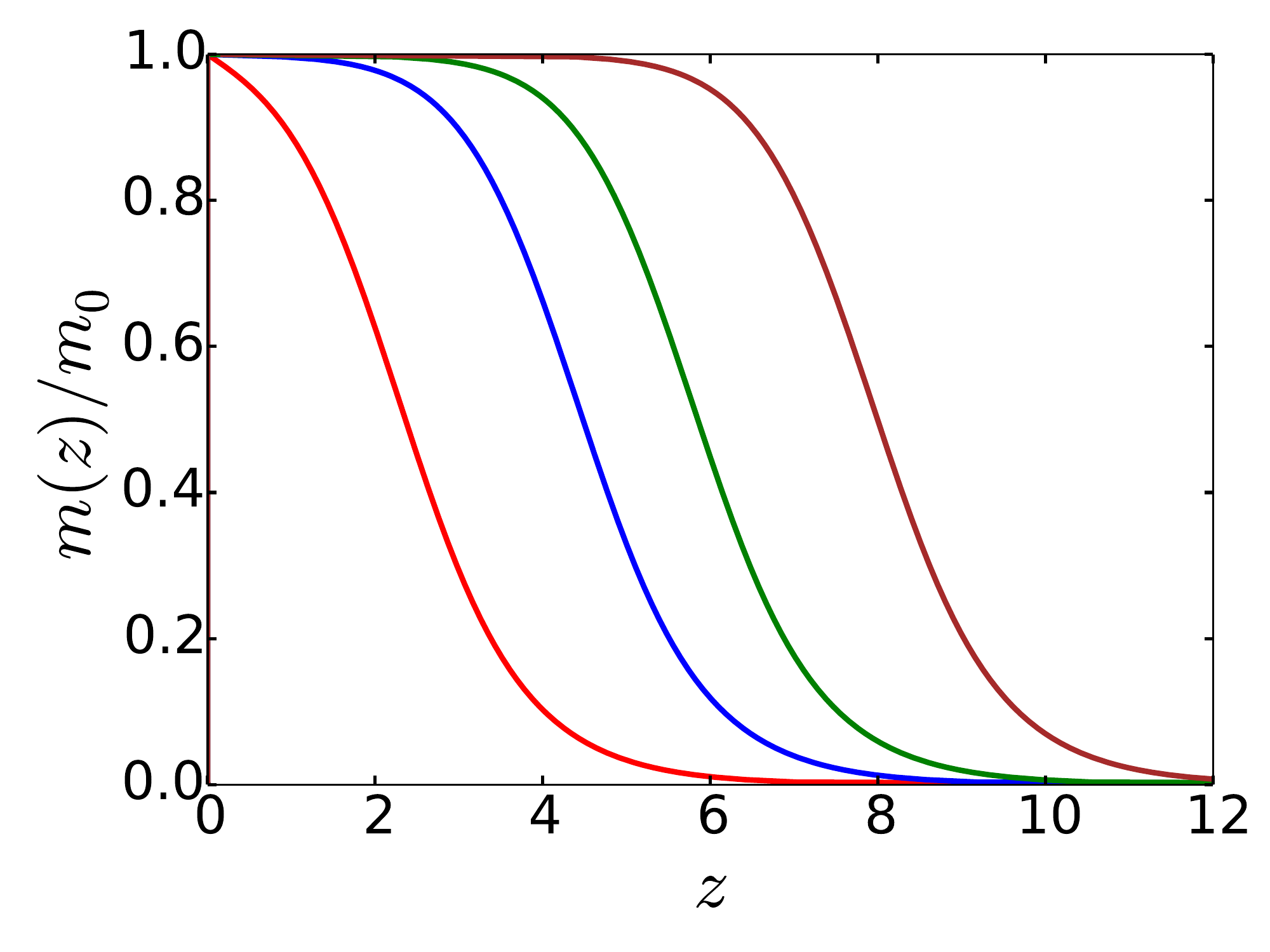}
\caption{\label{SIO} Magnetization $m(z)$, in units of the surface value, as function of the distance from the surface $z$ 
for different values of $h$ in the case of an ordered wetting layer, namely within the coexistence region where the stable phase is disordered and the ordered phase is metastable. We note the growing thickness of 
the ordered region as the first order critical point at $h_c\simeq 2.63113$ is approached.}
\end{figure}

In what follows we shall assume 
Hamiltonian parameters such that the model is within the coexistence region. As mentioned, we choose $m_\infty$ as the equilibrium value, which thus corresponds to the 
absolute minimum of the potential, $V(m)\geq V(m_\infty)=V_\infty$, $\forall m$. For convenience we  
shift the energy by $-V_\infty$, so that the potential turns into 
\be
V(m)\to V(m)-V_\infty \geq 0.\label{shifted-V(m)}
\ee
On the other hand, the surface value $m_0$ is chosen as the metastable minimum, with potential energy 
$V(m_0)-V(m_\infty)\equiv \Delta V_0\geq 0$. 

We first study the saddle point configuration that minimises the energy functional Eq.~\eqn{continuum:model}. The minimum is obtained when the conjugate momentum is zero, i.e. $\phi(z)=0$, and $m(z)$ satisfies the differential equation:
\begin{eqnarray}\label{func:der}
\frac{\delta \mathcal{H}}{\delta m} &=& \frac{dV(m)}{dm}- K_\perp(m)\,m''\nonumber\\
&& -\fract{1}{2}\,K'_\perp(m)\,m'\,^2 =0 ,\label{diff-eq}
\end{eqnarray}
with fixed boundary conditions $m(0)=m_0$ and $m(\infty)=m_\infty$. Eq.~\eqn{func:der} admits an integral of motion:
\begin{eqnarray}
I &=& \fract{K_\perp\big(m(z)\big)}{2}\;m'(z)^2 - V\big(m(z)\big),\label{eq-I}
\end{eqnarray}
and corresponds to the energy of a particle in the inverted potential $-V(m)$.~\cite{Coleman} 
The lowest energy trajectory is that one with lowest $I$, which, because  
$I\geq  -V\big(m(z)\big)$, $\forall z$, and through the definition 
Eq.~\eqn{shifted-V(m)}, must correspond to $I=0$. 
We thus obtain an implicit formula for $m(z)$:
\begin{eqnarray}\label{implicit:integral}
\int_{m_0}^{m(z)} dm \;\sqrt{\fract{K_\perp(m)}{2\,V(m)}}  = \pm z,
\end{eqnarray}
where the $+$ refers to the case $m_0<m_\infty$, and the $-$ to the opposite case. 

We thus have all elements required to study the two different, but specular phenomena: 
\begin{itemize}
\item{\sl Ordered wetting layer --} the 
bulk is disordered, $m_\infty = 0$, but the surface is constrained to be ordered, $m_0\not =0$. This case should be relevant to an interface between a metal and a Mott insulator. In section 
\ref{A model for a metal-Mott insulator interface} we shall discuss a better modeling of such situation. 
\item{\sl Disordered wetting layer --} This is the opposite case of an ordered bulk, $m_\infty\not =0$, and the surface forced to be disordered, $m_0 = 0$. This should be representative of an interface between the vacuum and a correlated metal.  
\end{itemize}

\subsection{Wetting }
\label{Wetting phenomenon}

\begin{figure}[bht!]
\includegraphics[width = 0.99\columnwidth]{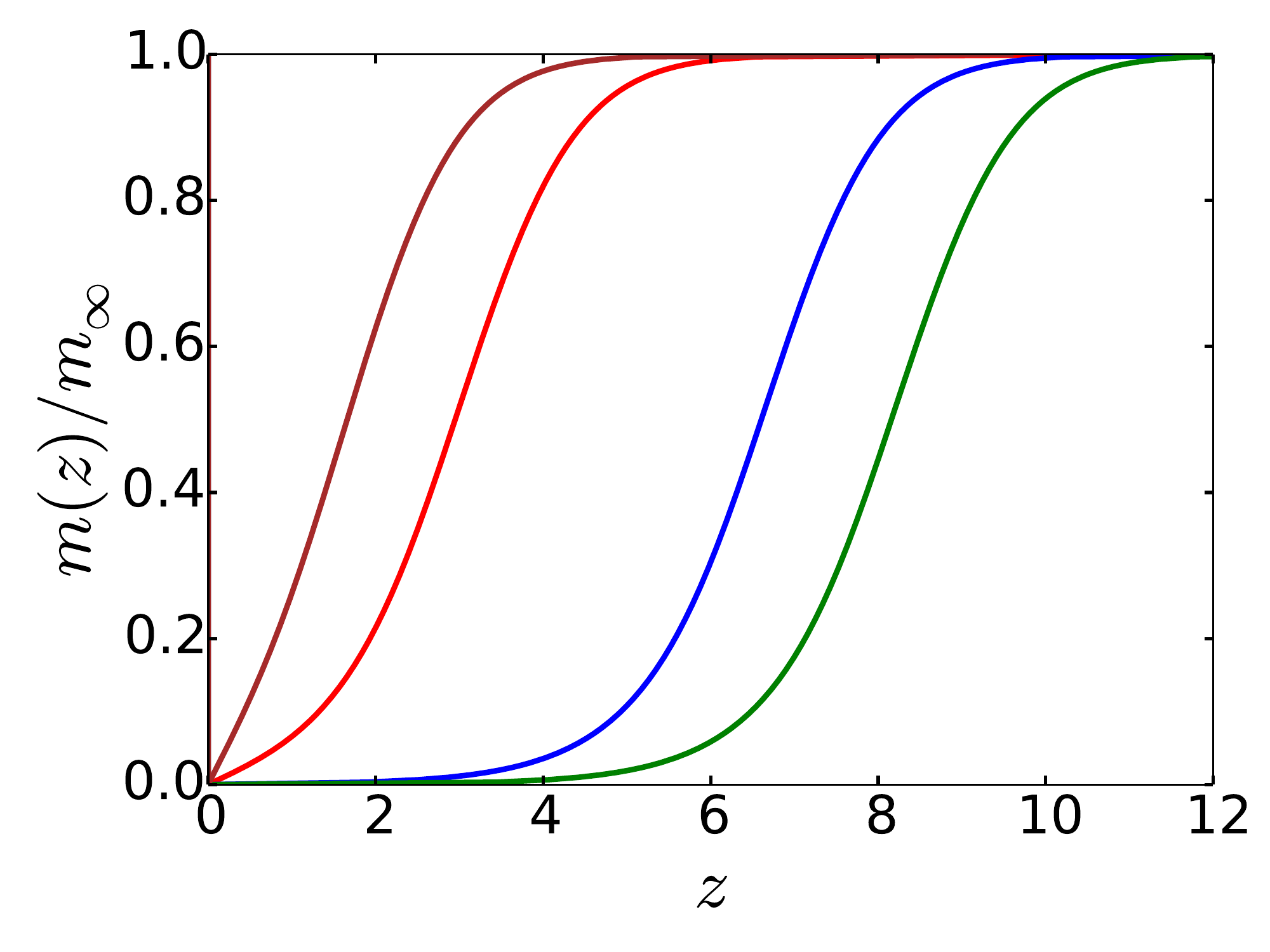}
\caption{\label{SID} Magnetization $m(z)$, in units of the bulk value, as function of the distance $z$  from the surface for different values of $h$ in the case of a disordered wetting layer, namely within the coexistence region where the stable phase is ordered and the disordered phase is metastable. Note 
the growing thickness of 
the disordered region as the first order 
transition 
point at $h_c\simeq 2.63113$ is approached.}
\end{figure}

In Fig.~\ref{SIO} and Fig.~\ref{SID} we show the numerical solution of Eq.~\eqn{implicit:integral} in both cases of either ordered or of disordered wetting layers, respectively, for the model of Eq.~\eqn{continuum:model} with parameters $\gamma  = \tilde{\gamma}=\gamma_\perp= \tilde{\gamma}_\perp=1/2$, in which case the coexistence region 
is $2\leq h \leq \sqrt{8}\simeq 2.83$,  
with the first order transition 
at
$h=h_c\simeq 2.63113$. In both cases 
a wetting layer forms at the surface,  its thickness growing as the first order quantum phase transition is approached at $h_c$.

The critical properties of  classical wetting 
at an interface have been extensively studied over many years, see e.g. 
Refs.~\onlinecite{Cahn1977,Lipowsky1982,Brezin&Halperin1983,Fisher&Huse1985,deGennes1985,Lipowsky&Fisher1987,Jin&Fisher1993,Bonn&Ross2001,Tartaglino2005,Parry&Rascon2008,Bonn2009,Parry&Rascon2009,Indekeu2010,Rath&Spivak2011,Jakubczyk2012}, which is not at all an exhaustive list. Here we will 
rederive some simple results in order to set up the terminology for the subsequent discussion 
of
quantum fluctuations applicable to our case. 

\begin{figure}[bht!]
\includegraphics[width = 0.99\columnwidth]{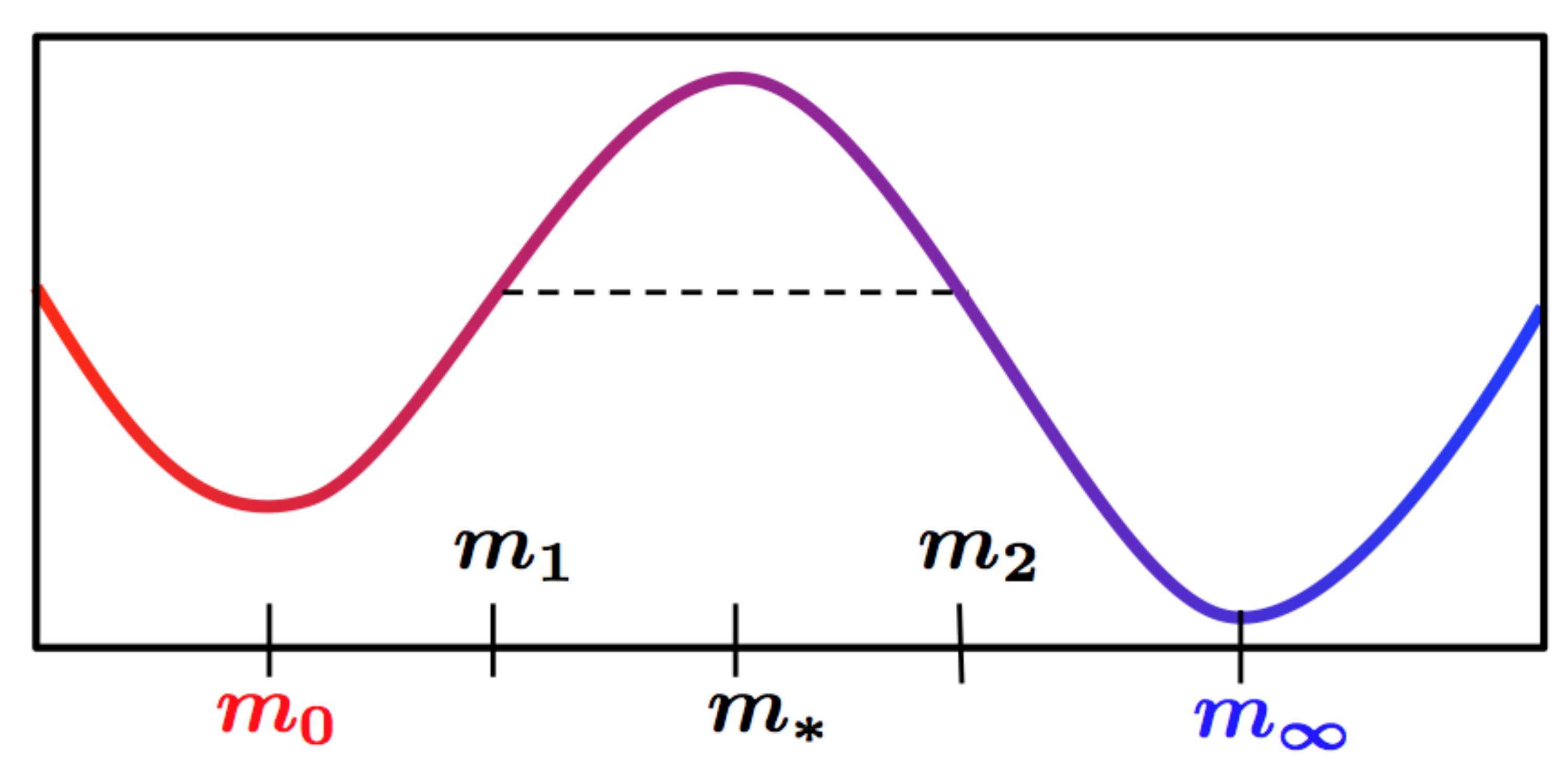}
\caption{\label{mvalues} Values of the magnetization relevant to define the wetting layer. We assume that the wetting interface has a width that corresponds to the distance required to go from the two inflection points $m_1$ and $m_2$. The wetting layer is defined as the distance to reach $m_1$ starting from the surface value $m_0$.}
\end{figure}

We assume 
the wetting interface to be located at $z=z_*$ when $m(z)$ reaches the top of 
the barrier that separates the two minima of $V(m)$, see Fig.~\ref{slab}.  Moreover, we assume that the width $W$ 
of the interface is the distance needed to go from the inflection point $m_1$ at the left of $m_*$ to that at the 
right, $m_2$, see Fig.~\ref{mvalues} where  for simplicity we take $m_0<m_\infty$, i.e. the case of a disordered wetting layer. Through Eq.~\eqn{implicit:integral} we find that 
\be
W = \int_{m_1}^{m_2} dm \;\sqrt{\fract{K_\perp(m)}{2\,V(m)}}.\label{width}
\ee

The thickness of the wetting layer can be 
defined as the distance required to reach $m_1$ from the surface value $m_0$, i.e. 
\be
l = \int_{m_0}^{m_1} dm \;\sqrt{\fract{K_\perp(m)}{2\,V(m)}}.\label{l}
\ee
Close to the first order transition, $h\to h_c$, the integral is dominated by the region close to $m_0$. If we 
approximate 
\[
V(m) \simeq \Delta V_0 + \fract{\kappa_0}{2}\,\big(m-m_0\big)^2,
\]
where $\kappa_0>0$, $K_\perp(m)\simeq K_\perp(m_0)$,  and $\Delta V_0 >0$ vanishing linearly when $h\to h_c$, we find 
\bea
l &\simeq& \sqrt{\fract{K_\perp(m_0)}{\kappa_0}}\; \sinh^{-1} \sqrt{\fract{\kappa_0}{2\Delta V_0}}\;
\big(m_1-m_0\big)\nonumber\\
&\simeq& \sqrt{\fract{K_\perp(m_0)}{\kappa_0}}\; \ln \left(2\;\sqrt{\fract{\kappa_0}{2\Delta V_0}}\;
\big(m_1-m_0\big)\right)\nonumber\\
&\simeq& - \fract{1}{2}\sqrt{\fract{K_\perp(m_0)}{\kappa_0}}\; \ln \big|h-h_c\big|.\label{xi_perp}
\eea
The thickness of the wetting layer diverges logarithmically as the first order bulk transition is 
approached, $\Delta V_0\to 0$, thus displaying a surface critical phenomenon despite the discontinuous nature of the bulk transition.~\cite{Cahn1977,Lipowsky1982} Within the same approximation and for 
$z < l$, 
\be
m(z) \simeq m_0 + \sqrt{\fract{2\Delta V_0}{\kappa_0}}\;\sinh \sqrt{\fract{\kappa_0}{K_\perp(m_0)}}\;
z.\label{m(z)-z small}
\ee

On the contrary, when $z\gg l$, hence $m(z)\simeq m_\infty$, we can approximate
\[
V(m) \simeq \fract{\kappa_\infty}{2}\,\big(m-m_\infty\big)^2,
\]
so that 
\be
m_\infty - m(z) \sim \exp\bigg(-\sqrt{\fract{\kappa_\infty}{K_\perp(m_\infty)}}\; z\bigg). 
\label{m(z)-z big}
\ee

We observe that 
\bea
\xi_0 &=& \sqrt{\fract{K_\perp(m_0)}{\kappa_0}},\label{xi_0}\\
\xi_\infty &=& \sqrt{\fract{K_\perp(m_\infty)}{\kappa_\infty}},\label{xi_infty}
\eea
are precisely 
the bulk correlation lengths in the metastable and stable phases, respectively, 
so that 
\be
l \simeq \xi_0\;\ln\big|h-h_c\big|.\label{l_new}
\ee


\subsection{A model for a metal-Mott insulator interface}
\label{A model for a metal-Mott insulator interface}

If we aim to model a metal-Mott insulator interface, where, as mentioned, 
the Mott insulator and the metal translate into the paramagnetic and ferromagnetic phases of the Hamiltonian Eq.~\eqn{continuum:model}, respectively, it is 
convenient 
to consider a slightly different interface problem, 
introduced by Cahn back in 1977~\cite{Cahn1977} and by now the paradigm of classical wetting transitions. 

\begin{figure}[thb!]
\includegraphics[width = 0.9\columnwidth]{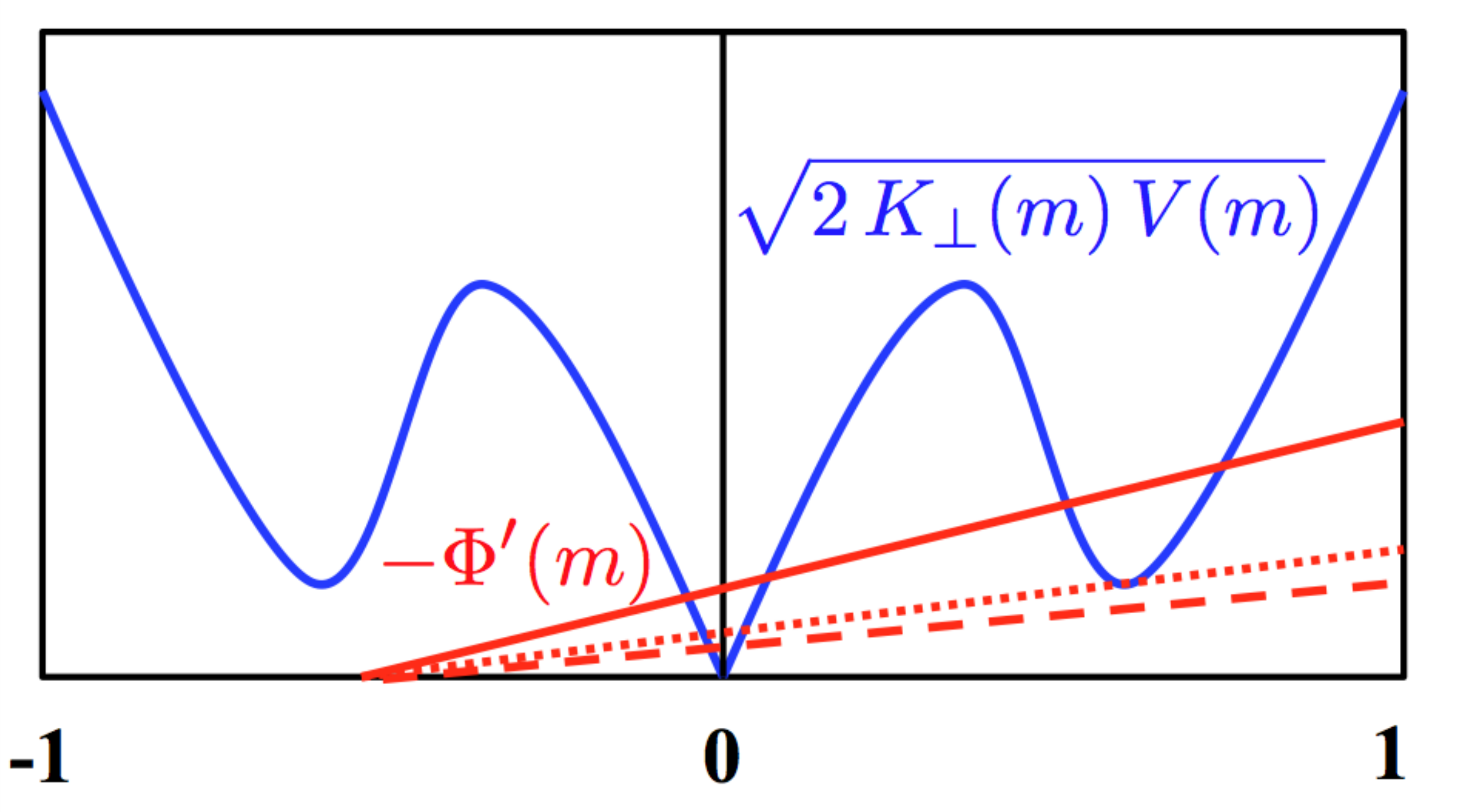}
\caption{\label{Phi-fig} Graphical solution of Eq.~\eqn{interface-m_0}. The blue curve is $\sqrt{2\,K_\perp(m_s)\,V(m_s)}$, while 
the red straight lines 
represent
$- \Phi'(m_s)$ for increasing values of 
$\lambda_\perp$, from bottom to top.}
\end{figure}

\begin{figure}[bht!]
\includegraphics[width = 0.8\columnwidth]{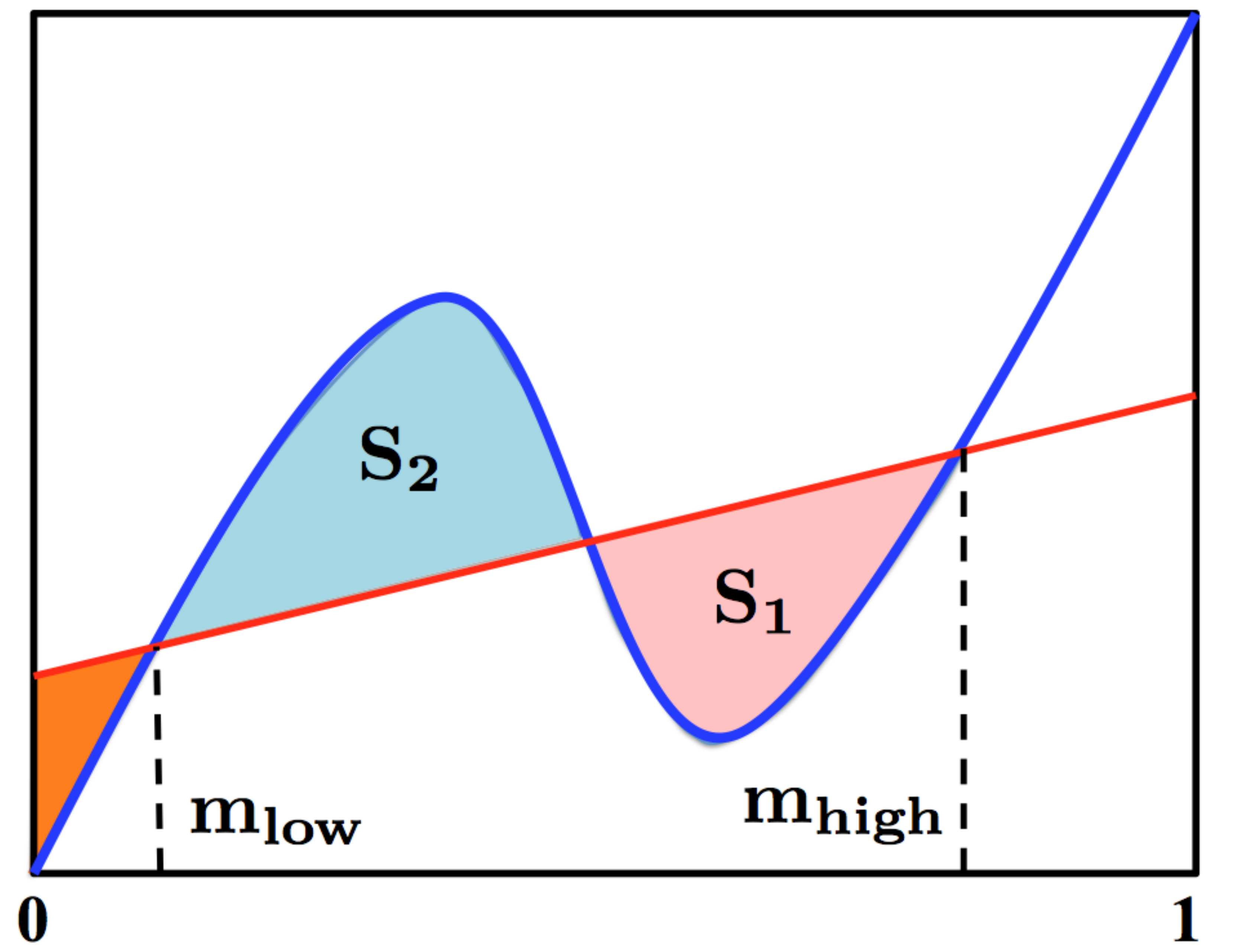}
\caption{\label{S1-S2} The two minima $m_\text{low}$ and 
$m_\text{high}$ for the surface magnetisation $m_s$ above the threshold value of the coupling to the ferromagnet. The total energy is given by the shaded areas where the pink and orange ones are taken with negative sign and the cyan one with positive sign. }
\end{figure}

In that model, the paramagnetic, i.e. Mott-insulating, slab is 
still described by Eq.~\eqn{continuum:model} with a stable minimum at $m_\infty=0$, but one assumes the surface layer at $z=0$ to be coupled to a very stiff ferromagnet, i.e. a very good metal, whose magnetisation $m_\text{metal}\simeq 1$ is therefore not affected by the slab. The coupling is thus modelled by 
the surface potential~\cite{Cahn1977} 
\be
\Phi\big(m_s\big) = -\lambda_\perp\,\Big(m_s\,m_\text{metal} 
+m_s^2\,m_\text{metal}^2\Big),\label{Phi}
\ee   
where we denote $m(0)=m_s$ and, as mentioned, $m_\text{metal}\simeq 1$ is constant. 
As before we start by the saddle point configuration where the conjugate field $\phi(z)=0$ and $m(z)$ is the solution of 
Eq.~\eqn{func:der}, or, through Eq.~\eqn{eq-I} with $I=0$, of 
\be
m'(z) = - \sqrt{\fract{2V\big(m(z)\big)}{K_\perp\big(m(z)\big)}}
\;,\label{interface-m}
\ee
where we now choose the minus sign since $m(z)$ decreases with $z$. 
The difference 
is that the surface magnetisation $m(0)=m_s$, i.e. the boundary condition of Eq.~\eqn{interface-m}, must 
now
be determined by minimising the total energy 
\bea
E\big(m_s\big) &=& \Phi\big(m_s\big) 
+ \int_0^\infty \!\!dz \bigg[ \fract{K_\perp\big(m(z)\big)}{2}\,\left(\fract{\partial m(z)}{\partial z}\right)^2 \nonumber\\
&&  + V\big(m(z)\big)\bigg] = 
\Phi\big(m_s\big) 
+ 2\,\int_0^\infty \!\!dz V\big(m(z)\big) \nonumber\\
&=& \Phi\big(m_s\big) 
+ \int_{0}^{m_s} \!\!dm\; \sqrt{2\,K_\perp(m)\,V(m)}
\label{interface-E}\\
&=& \int_{0}^{m_s} \!\!dm\,\bigg[
\sqrt{2\,K_\perp(m)\,V(m)} - \Big(-\Phi'(m)\Big)\bigg],
\nonumber
\eea
where the last equation holds since $\Phi(0)=0$. The surface value 
$m_s$ thus satisfies the equation 
\be
\sqrt{2\,K_\perp(m_s)\,V(m_s)} = - \Phi'(m_s),
\label{interface-m_0}
\ee
which is graphically shown in Fig.~\ref{Phi-fig} for increasing 
values of $\lambda_\perp>0$.  
We observe that for small $\lambda_\perp$, dashed line in the figure, there are only two crossing 
points, at the left and at the right of $m=0$, 
the latter 
a
minimum of the energy and the former a maximum. Only above a threshold, shown as 
a
dotted line,  
two other extrema appear. 
In that case there are two minima, denoted as $m_\text{low}$ and $m_\text{high}$ in 
Fig.~\ref{S1-S2}.
Through the last expression in Eq.~\eqn{interface-E}, we readily obtain that the energy of the solution $m_s=m_\text{low}$ is minus the 
orange shaded area shown in Fig.~\ref{S1-S2}, while that of 
$m_s=m_\text{high}$ is the cyan shaded area, $S_2$ in the figure, minus 
the sum of the orange plus the pink, $S_1$ in the figure, areas. 
In other words 
\be
E\big(m_\text{low}\big) - E\big(m_\text{high}\big) = S_1 - S_2 \equiv S,\label{interface-S}
\ee
where $S$ is the so-called {\sl spreading coefficient}, see e.g. 
Ref.~\onlinecite{deGennes1985}. 
It follows that, if $S<0$, the absolute minimum is obtained 
for $m_s=m_\text{low}$, which corresponds to what is 
called  
{\sl partial wetting}.~\cite{deGennes1985} On the contrary, when 
$S>0$ the absolute minimum has surface magnetisation 
$m_s=m_\text{high}$ 
which
corresponds to {\sl complete wetting}. $S=0$ 
signals a first order transition between the two solutions, 
where
the surface magnetisation 
jumps discontinuously from $m_\text{low}$ to $m_\text{high}$.

If we still denote as $m_0<m_\text{high}$ the metastable ferromagnetic, i.e. metallic, minimum of $V(m)$, the thickness of the wetting layer increases with respect to $l$, 
Eq.~\eqn{xi_perp}, i.e. to the situation discussed in section~\ref{Wetting phenomenon}, by 
\be
\delta l = \int_{m_0}^{m_\text{high}}\, dm\, 
\sqrt{\fract{K_\perp(m)}{2V(m)}}\;,
\ee      
which also diverges logarithmically as the bulk transition is approached. 

\subsection{Fluctuations in the harmonic approximation}

The distiguishing feature of quantum wetting from its classical counterpart is represented by quantum fluctuations. 
We therefore add quantum fluctuations to the saddle point solution 
discussed in section~\ref{Wetting phenomenon} within the harmonic approximation. We start from the Hamiltonian:
\bea
\mathcal{H} &=& \int_0^L \!\!dz \Bigg[ \fract{K_\perp\big(m(z)\big)}{2}\,\left(\fract{\partial m(z)}{\partial z}\right)^2 
+ V\big(m(z)\big)\nonumber\\
&& \phantom{\int_0^L \!\!dz \Bigg[ } + 2h\,\sqrt{1-m^2(z)}\;\sin^2 \phi(z)\Bigg]
,\label{E(m)}
\eea
this time defined within the interval $z\in [0,L]$, where $L$ will be sent to infinity afterwards.  
The functional Eq.~\eqn{E(m)} is minimized by $\phi=0$ and by the solution of Eq.~\eqn{eq-I} with $I=0$, which we shall denote as $m_\text{sp}(z)$, with boundary conditions $m_\text{sp}(0)=m_0$ and 
$m_\text{sp}(L\to\infty)=m_\infty$. 
We perform a change of variable $m(z)\to q(z)$ where 
\be
\sqrt{K_\perp(m)}\;d m = d q,
\ee
so that the saddle point trajectory $q_\text{sp}(z)$ satisfies 
\bea
q''_\text{sp}(z) &=& \fract{\partial V\big(m(q)\big)}{\partial q}_{\big|q=q_\text{sp}(z)}
\label{qddot}\\
&=& \left(\fract{1}{\sqrt{K_\perp(m)}}\;\fract{\partial V(m)}{\partial m}\right)
_{\big|m=m\big(q_\text{sp}(z)\big)},\nonumber\\
\fract{q'_\text{sp}(z)^2}{2} &=& V\Big(m\big(q_\text{sp}(z)\big)\Big),\label{qdot}
\eea
from which it follows that $q''_\text{sp}(0)=q''_\text{sp}(L)=0$. Moreover, because the potential has been rigidly shifted in such a way that the absolute  minimum at $z=L$ is zero, it also follows that $q'_\text{sp}(L)=0$.

If we vary $q(z)$ at fixed boundary conditions, i.e. 
\be
q(z) = q_\text{sp}(z) + \psi(z),
\ee
with small $\psi(z)$ such that $\psi(0)=\psi(L)=0$, which corresponds to 
\[
m(z) = m_\text{sp}(z) + \sqrt{\fract{1}{K_\perp\big(m_\text{sp}(z)\big)}}\; \psi(z),
\]
the energy functional in Eq.~\eqn{E(m)},  up to second order  in $\psi(z)$ and $\phi(z)$, changes into 
\bea
\mathcal{H} &\simeq&  E\big[m_\text{sp}(z)\big]  + \fract{1}{2}\,\int_0^\infty dz\,
\Bigg[ \Big(\partial\psi(z)\Big)^2 \nonumber\\
&& + U^{(2)}(z)\,\psi(z)^2 + 4h\,\sqrt{1-m_\text{sp}^2(z)\,}\;\phi(z)^2
\Bigg],\qquad \label{H-fluct}
\eea
where
\be
U^{(n)}(z) = \fract{\partial^n V\big(m(q)\big)}{\partial q^n}_{\big|q=q_\text{sp}(z)}.\label{U(z)}
\ee 
The potential $U^{(2)}(z)$ has the following limiting values 
\ba
U^{(2)}(z\to 0) &\to&  \fract{\kappa_0}{K_\perp(m_0)} = \xi_0^{-2} > 0,\\
U^{(2)}(z\to L) &\to&  \fract{\kappa_\infty}{K_\perp(m_\infty)} = \xi_\infty^{-2} > 0,
\ea
and has a negative minimum $-U_* < 0$ approximately at $z=z_*$; in other words it describes a potential well centered around the wetting interface.  

\subsubsection{An auxiliary eigenvalue problem}
Let us consider the auxiliary problem~\cite{Coleman-book,Brezin&Halperin1983-2} that consists of solving the following Schr{\oe}dinger eigenvalue equation in $z\in[-L,L]$ with vanishing conditions at $z=\pm L$: 
\be
\bigg(-\fract{\partial^2}{\partial z^2} + U(z)\bigg)\,\psi_n(z) = \epsilon_n\,\psi_n(z),\label{Schroedinger}
\ee
where
 \begin{equation}
 U(z) = \left\{
 \begin{array}{cc}
 \displaystyle U^{(2)}(z) & \displaystyle \mbox{when } z \geq 0 \\ 
\displaystyle U^{(2)}(-z) &\displaystyle \mbox{when } z <0
\end{array}
 \right..
 \end{equation}
 The  potential $U(z)$ thus describes two symmetric wells centered roughly at $\pm z_*$. 
The eigenstates of Eq.~\eqn{Schroedinger} can be always chosen real and distinguished into even, 
$\psi^\text{even}_n(z)=\psi^\text{even}_n(-z)$, and odd, $\psi^\text{odd}_n(z)=-\psi^\text{odd}_n(-z)$, 
with eigenvalues $\epsilon^\text{even}_n$ and $\epsilon^\text{odd}_n$, respectively, with $n\geq 0$. 
By continuity of the wavefunction and its first derivative at the origin, 
$\partial\psi^\text{even}_n(0) = 0$ 
and 
$\psi^\text{odd}_n(0) = 0$, while, by construction, $\psi^\text{even}_n(\pm L) =\psi^\text{odd}_n(\pm L) = 0$. We now note that the even-parity wavefunction defined as $\psi(z)=q'_\text{sp}(z)$ for $z>0$, 
and $\psi(z)=q'_\text{sp}(-z)$ for $z<0$, 
is a solution of Eq.~\eqn{Schroedinger} with 
vanishing eigenvalue that satisfies all appropriate boundary conditions. Since it is nodeless, it is actually the ground state of Eq.~\eqn{Schroedinger}: $\psi(z)\equiv \psi^\text{even}_0(z)$ with eigenvalue 
$\epsilon^\text{even}_0=0$. It thus follows that, since all eigenvalues in the odd parity sector must be necessarily greater than the ground state energy, then $\epsilon^\text{odd}_n > 0$, $\forall n\geq 0$; 
the saddle point is stable to fluctuations. 

Now suppose the system is very close to the first order transition, i.e. $\Delta V_0\ll 1$. In this case the two 
potential wells in $U(z)$ are very far apart and we can safely assume that the ground state 
$\psi^\text{even}_0(z)$, with energy $\epsilon^\text{even}_0=0$, is the symmetric combination of the ground state wavefunctions of each well, while the first excited state is the antisymmetric combination, hence is the lowest energy state in the odd sector. The two states are split by the quantum tunneling amplitude between the two wells. Within the WKB approximation we can estimate that energy splitting as 
\bea
\epsilon^\text{odd}_0 - \epsilon^\text{even}_0 &=& \epsilon^\text{odd}_0 \simeq  2\,\fract{\sqrt{U_*}}{W}\;
\text{e}^{-2\,l/\xi_0} \nonumber\\
&\simeq& 2\,\fract{\sqrt{U_*}}{W}\;\big|h-h_c\big|\,
,\label{split}
\eea
which is therefore the leading term of the lowest eigenvalue in the odd sector. 
\begin{figure}[htb!]
\includegraphics[width = 0.99\columnwidth]{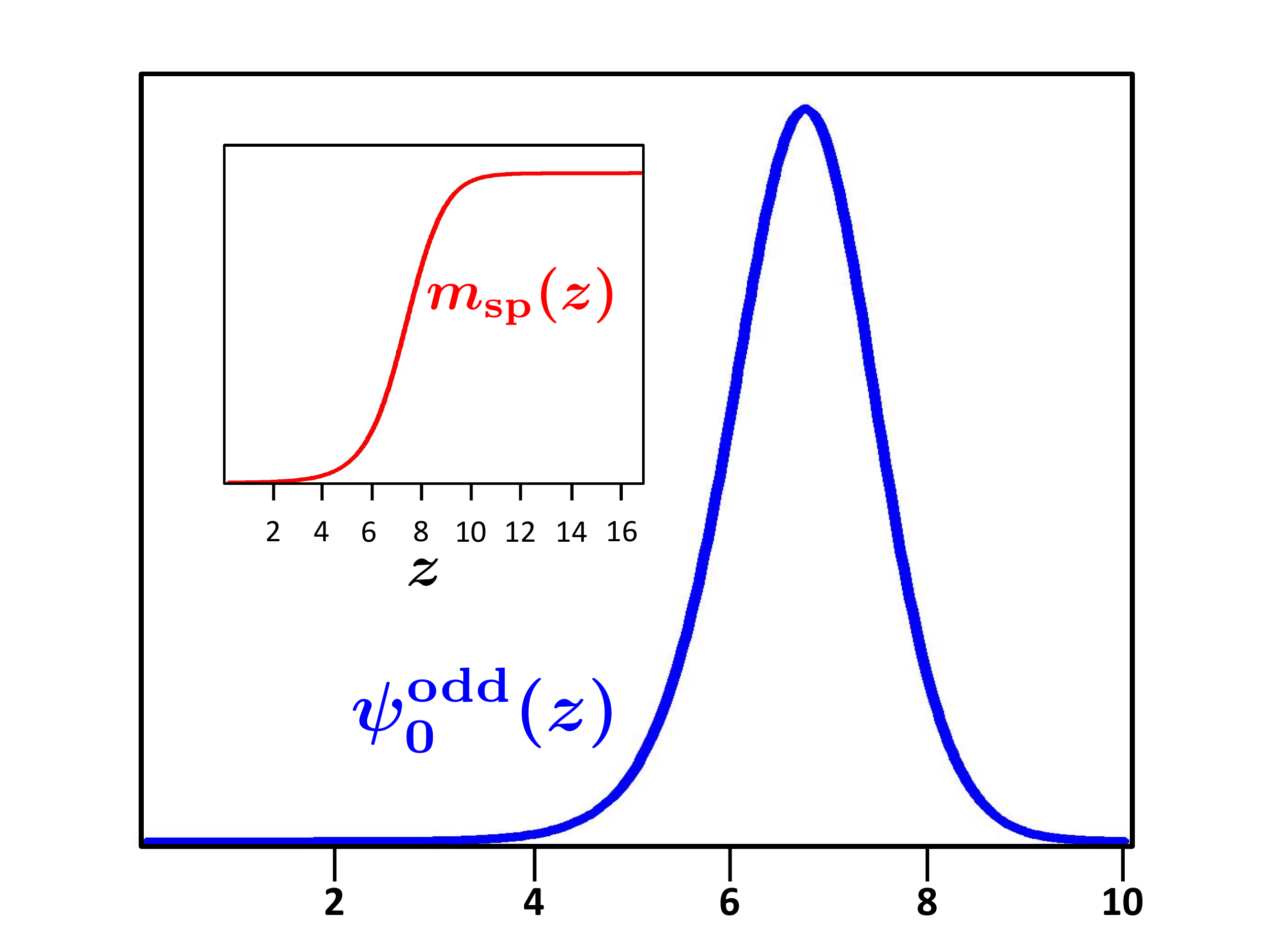}
\caption{\label{psi&m} The lowest energy wavefunction $\psi^\text{odd}_0(z)$ in the odd sector for 
$h=2.6311298<h_c$ very close to the transition, $\Delta V_0 = 2.3\,10^{-7}$. In the inset we show the saddle point magnetization profile $m_\text{sp}(z)$.
}
\end{figure}
In Fig.~\ref{psi&m} we plot $\psi^\text{odd}_0(z)$ in the case of a disordered wetting layer very close to the first order transition, when $\Delta V_0 = 2.3\,10^{-7}$. 

We believe that, besides the above two states, with energies $\epsilon^\text{even}_0$ 
and $\epsilon^\text{odd}_0$, there are no other eigenstates that are bound within the potential wells. In other words, all other eigenstates have energy greater than $\text{min}\big(\xi_0^{-2},\xi_\infty^{-2}\big)$ and are not localized on the wetting interface.  We have no rigorous proof of the above statement, even though we find it is confirmed by the semiclassical WKB quantization rule. 

To conclude, we note that the existence of a zero energy mode of the Hamiltonian 
Eq.~\eqn{Schroedinger} could be anticipated. Indeed, if we consider the magnetisation profile $m_\text{sp}(z-\epsilon)$, with $\epsilon\ll 1$, which corresponds to $\delta q(z) \simeq 
-\epsilon\,q'_\text{sp}(z) = -\epsilon\,\psi^\text{even}_0(z)$, i.e. just to the zero-energy mode, the energy changes into 
\[
\mathcal{H}(\epsilon) = \int_{-\epsilon}^\infty dz\,
2\,V\big(m_\text{sp}(z)\big),
\]
whose second derivative with respect to $\epsilon$ indeed vanishes
\[
\fract{\partial^2 \mathcal{H}(\epsilon)}{\delta\epsilon^2} 
= - 2 \fract{\partial V\big(m_\text{sp}(z)\big)}{\partial z}
_{\big|z=0} = 0.
\] 

\subsubsection{Explicit expression of quantum fluctuations}

We now expand $\psi(z)$ and $\phi(z)$ in Eq.~\eqn{H-fluct} in a complete basis  
within the interval $z\in[0,L]$ with vanishing conditions, which we can take as the set of odd eigenstates 
of Eq.~\eqn{Schroedinger}, 
\bea
m(z) &=& m_\text{sp}(z) + \fract{1}{\sqrt{K_\perp(m_\text{sp}(z))}}\; \psi(z) \label{psi-x}\\
&=& m_\text{sp}(z) +
\sqrt{\fract{2}{K_\perp(m_\text{sp}(z))}}\; \sum_{n}\,x_n\,\psi^\text{odd}_n(z),\nonumber\\
\phi(z) &=& \sqrt{2\,K_\perp(m_\text{sp}(z))}\;\sum_{n}\,p_n\,\psi^\text{odd}_n(z),\label{phi-x}
\eea
where $[x_n,p_m]=i\,\delta_{nm}$ so that 
\[
\big[m(z),\phi(z')\big] = i\,\delta(z-z'),
\]
are indeed conjugate variables. It follows that  the Hamiltonian Eq.~\eqn{H-fluct} becomes
\bea
&&\mathcal{H}- E\big[m_\text{sp}(z)\big] =   + \frac{1}{2}\,\sum_n\, \epsilon^\text{odd}_n\,
x_n^2 + \gamma^\text{odd}_{nn}\,p_n^2 \nonumber\\
&& \qquad + \frac{1}{2}\,\sum_{n\not= m}\,\gamma^\text{odd}_{nm}\,p_n\,p_m,\label{H-fluct-1}
\eea
where 
\ba
\gamma^\text{odd}_{nm} &=& 8h\,\int_0^L dz\, \bigg[K_\perp\big(m_\text{sp}(z)\big)\,
\\
&&\phantom{8h\,\int_0^L dz\;\;\;\;}\,\sqrt{1-m_\text{sp}(z)^2\,}\; 
\psi_n^\text{odd}(z)\psi_m^\text{odd}(z)\bigg].
\ea
Since $\psi^\text{odd}_0(z)$ is localized on the interface, while $\psi^\text{odd}_n(z)$ for $n>0$ 
are extended at least over the whole wetting thickness $l$, 
the coupling $\gamma^\text{odd}_{0n}\sim l^{-1/2}$  vanishes approaching the transition. We can 
therefore treat it within perturbation theory and obtain for the $n=0$ mode the effective Hamiltonian 
\bea
\mathcal{H}_0 &=& \fract{\epsilon^\text{odd}_0}{2}\;x_0^2 + \fract{\gamma_0}{2}\;p_0^2 = 
\omega_0\,\Big(b^\dagger_0\,b^\dagga_0 + \frac{1}{2}\Big),\label{H-mode=0}
\eea
where 
\[
\gamma_0 \simeq \gamma^\text{odd}_{00} + \sum_{n,m>0}\, \gamma^\text{odd}_{0n}\; 
\left(\fract{1}{\hat{\gamma}}\right)^{-1}_{nm}\,\gamma^\text{odd}_{m0},
\]
with $\hat{\gamma}$ the symmetric real matrix with elements $\gamma^\text{odd}_{nm}$, 
and $b_0$ the bosonic operator that diagonalize the Hamiltonian \eqn{H-mode=0} with eigenvalue 
\be
\omega_0 = \sqrt{\epsilon^\text{odd}_0\,\gamma_0\,}\, \simeq 
\big|h-h_c\big|^{1/2},\label{eigenvalues}
\ee
which vanishes at the first order point, $h=h_c$, when the interface can be translated at no energy cost. 

Close to the transition, the static fluctuations of $q(z)$ with respect to the saddle point profile $q_\text{sp}(z)$ is dominated by the $n=0$ mode, 
\be
\big\langle \,\Big(q(z)-q_\text{sp}(z)\Big)^2\,\big\rangle \simeq \fract{\gamma_0}{\omega_0}
\; \psi^\text{odd}_0(z)^2 \sim \big|h-h_c\big|^{-1/2},\label{fluctuations-q(z)}
\ee
and grow when $h\to h_c$, unlike the fluctuations of the conjugate variable $\phi(z)$. This behaviour simply reflects the unbinding of the wetting interface, whose detailed description would require going beyond the harmonic approximation. We observe that the square root divergence in Eq.~\eqn{fluctuations-q(z)} differs from the classical result~\cite{Lipowsky-ZPB-1984}, which, through Eq.~\eqn{H-mode=0}, predicts a more singular behaviour $\sim 1/\epsilon^\text{odd}_0 \sim \big|h-h_c\big|^{-1}$.  Therefore, even though the saddle point has evidently the same critical properties as in the classical case, the quantum fluctuations behave differently from the classical ones.

\subsubsection{Back to the metal-Mott insulator interface}

In the case of the metal-Mott insulator interface modelled in 
section~\ref{A model for a metal-Mott insulator interface} 
through a paramagnet slab supplied by an additional surface potential, the allowed eigenvectors $\psi^\text{odd}_n(z)$ 
(we keep using the label "odd", though now the wavefunction 
must not vanish anymore at $z=0$) still satisfy the eigenvalue equation \eqn{Schroedinger} for $z>0$ with 
the boundary condition 
\be
\fract{\partial \psi^\text{odd}_n(0)}{\partial z} 
= \fract{\partial^2 \Phi\big(m(q)\big)}{\partial q^2}_{\big|q=q_\text{sp}(0)}\,
\psi^\text{odd}_n(0),\label{interface-BC}
\ee 
where $\Phi(m)$ is defined by Eq.~\eqn{Phi}. 
We assume the model has complete wetting, i.e. $S>0$, see 
Eq.~\eqn{interface-S}. Since the second derivative of $\Phi(m)$ at the surface value $m_s=m_\text{high}$ is negative, the boundary condition 
Eq.~\eqn{interface-BC} actually corresponds to an attractive $\delta$-function potential 
\[
\delta U(z) = \delta(z)\,\fract{\partial^2 \Phi\big(m(q)\big)}{\partial q^2}_{\big|q=q_\text{sp}(0)} = - U_s\,\delta(0),
\]
with $U_s > 0$, that adds to $U(z)$ 
As before, the wavefunction 
$\psi(z)\propto q'_\text{sp}(z)$ is 
the 
solution of the Schr{\oe}dinger equation with vanishing eigenvalue, but with the "wrong" boundary condition 
\bea
\psi'(0) &=& q''_\text{sp}(0) = \fract{\partial V\big(m(q)\big)}{\partial q}_{\big|q=q_\text{sp}(0)}\label{wrong-BC}\\
&=& -\big(2K_\perp(m_s)V(m_s)\big)^{-1/2}\;\fract{\partial V(m)}{\partial m}_{\big|m=m_s}\;\psi(0),\nonumber
\eea
where we recall that 
\[
\sqrt{2K_\perp(m_s)V(m_s)} = - \Phi'(m_s).
\]
One easily realises that Eq.~\eqn{wrong-BC} corresponds to a more attractive $\delta$-function potential at $z=0$, which implies once again that the lowest allowed eigenvalue 
$\epsilon^\text{odd}_0$ is positive 
and vanishes only at the bulk first-order transition.  

\subsection{Capillary waves at the interface}
\label{Capillary waves at the interface}
So far we have assumed each layer to be infinitely connected, which is equivalent to a flat  
wetting interface. Let us now relax this assumption and consider the three dimensional Ginzburg-Landau Hamiltonian for the fluctuations, see also the Appendix, 
\bea
\mathcal{H} &\simeq&  E\big[m_\text{sp}(z)\big] +\fract{1}{2}\,\int d\br\, \int_0^\infty dz\,
\Bigg[ \Big(\partial\psi(z,\br)\Big)^2 \nonumber\\
&& + U(z)\,\psi(z,\br)^2 + 4h\,\sqrt{1-m_\text{sp}^2(z)\,}\;\phi(z,\br)^2\nonumber\\
&& + K_{||}\, \big(\boldsymbol{\nabla}\psi(z,\br)\big)^2
\Bigg],\label{H-fluct-3d}
\eea
where $\br$ is the coordinate in the plane of each layer and we added an intra-layer (longitudinal) stiffness term. Assuming translational 
symmetry within each layer, we can Fourier transform in $\br$ introducing the intra-layer, two-dimensional planar momentum $\bq$. 
If we now parametrize:
\bea
m(z,\bq) &=& m_\text{sp}(z) + \fract{1}{\sqrt{K_\perp(m_\text{sp}(z))}}\; \psi(z,\bq) \label{psi-x-q}\\
&=& m_\text{sp}(z) +
\sqrt{\fract{2}{K_\perp(m_\text{sp}(z))}}\; \sum_{n}\,x_{n\bq}\,\psi^\text{odd}_n(z),\nonumber\\
\phi(z,\bq) &=& \sqrt{2\,K_\perp(m_\text{sp}(z))}\;\sum_{n}\,p_{n,\bq}\,\psi^\text{odd}_n(z),\label{phi-x-q}
\eea
with the same wavefunctions as before and with 
\[
\big[x_{n,\bq},p_{m,-\bq'}\big] = i\,\delta_{nm}\,\delta_{\bq,\bq'},
\]
we find that the $n=0$ mode is described by the effective Hamiltonian 
\bea
\mathcal{H}_0 &=& \fract{1}{2}\sum_\bq\, \Big(\epsilon^\text{odd}_0
+ K_{||}\,q^2\Big)\,x_{0\bq}\,x_{0-\bq} + 
\gamma_0\,p_{0\bq}\,p_{0-\bq}\nonumber\\
&=& \sum_{\bq}\,\omega_{0\bq}\,
\Big(b^\dagger_{0\bq}\,b^\dagga_{0\bq} + \frac{1}{2}\Big),\label{H-mode=0-q}
\eea
where the eigenvalues 
\be
\omega_{0\bq} = \sqrt{\gamma_0\,\Big(\epsilon^\text{odd}_0 + K_{||}\,q^2
\Big)\,}, \label{eigenvalues-q}
\ee
allow defining a longitudinal correlation length 
\bea
\xi_{||}&\simeq& \sqrt{\fract{K_{||}}{\epsilon^\text{odd}_0}} \sim \big|h-h_c\big|^{-1/2},
\label{xi_||}
\eea
to be compared with $l$ in Eq.~\eqn{xi_perp}. We highlight that all other modes with $n>0$ are separated 
by a finite energy gap, so they cannot provide a finite damping at $T=0$ for the $n=0$ capillary wave.   
In the spin language, the wetting interface thus binds spin wave excitations that sink down below the gapped bulk continuum and propagate coherently along the interface.

The capillary wave contribution to the static fluctuations of the magnetisation $m(z,\br)$ with respect to the saddle point profile 
$m_\text{sp}(z)$
\be
\big\langle\; \Big(m(z,\br)-m_\text{sp}(z)\Big)^2\;\big\rangle \sim 
\int d\bq\; \fract{\gamma_0}{\;\omega_{0\bq}\;}\;,
\label{fluct-m(z,r)}
\ee
is not singular when $h\to h_c$, unlike in the classical finite temperature regime where it diverges logarithmically signalling the roughening of the wetting interface.

\section{Discussion and Conclusions}
\label{Conclusions}
We have studied some non-equilibrium as well as some non-homogeneous  properties of a quantum Ising model with a four-spin exchange, probably
the simplest representative of $Z_2$-invariant models that 
display a first-order quantum phase transition, i.e. a zero temperature symmetry-breaking transition upon varying a Hamiltonian parameter. 

First, we studied quantum quenches in the mean-field limit of infinite connectivity, where the corresponding Lipkin-Ising model can be solved exactly. In the coexistence region around the first order transition, we found that out-of-equilibrium the 
spatially homogeneous
system can remain trapped  into the metastable phase. 

Next, we studied the static but spatially inhomogeneous phenomenon that arises in the presence of an interface that pins the surface into the metastable phase. The latter, as in the classical wetting, extends over a finite length $l$ inside the bulk, which diverges as the first-order transition is approached. Specifically, within mean-field, which should provide the correct critical behavior in $d+1=4$ dimensions,~\cite{Brezin&Halperin1983-2} the thickness of the wetting layer diverges logarithmically at the 
first-order quantum phase transition, 
\be
l \sim -\ln\big|h-h_c\big|, \label{conclusion-xi_perp}
\ee
while the longitudinal fluctuations of the wetting interface are controlled by the correlation length
\be
\xi_{||} \sim \big|h-h_c\big|^{-1/2},\label{conclusion-xi_||}
\ee
and are associated to  
propagating modes with dispersion  
\be
\omega_{\bq} \propto \sqrt{ \xi_{||}^{-2}+q^2}\,,\label{conclusion-omega}
\ee
where $\bq$ is the momentum parallel to the interface. 
This branch of excitations localised on the wetting interface lies below the bulk spectrum, which has a finite gap in both ordered and disordered phases. It is therefore a coherent mode at low temperatures that cannot decay in the bulk continuum. At the transition it turns into an acoustic mode $\omega_\bq \sim q$ that at zero temperature does not lead to diverging fluctuations unlike the classical case at finite temperature. 

We hypothesize 
that the above results 
should be
applicable to quantum first-order Mott transitions,  
with some interesting consequences. The first is the possibility of nucleating long-lived droplets of a metastable metal inside a stable Mott insulator by external perturbations of finite-duration, which recalls  
recent experiments~\cite{WolfVO2-2014,Brazovskii2014} where signatures of non-thermal metallic phases have been observed after irradiation of Mott insulators by ultra-short laser pulses. 
Moreover, the observation that VO$_2$ electric double layer transistors formed at a solid/electrolyte interface show upon superficial charging a bulk-like metallization 
with an extension much beyond the electrostatic screening length,~\cite{Iwasa} is strongly suggestive of  a metallic wetting layer that forms at the interface. 

A more realistic modeling of interface phenomena close to a first order Mott transition must include long-range dispersion forces, 
ignored in the present treatment. As is well known,~\cite{Lipowsky-PRB1985,Dietrich1985,Dietrich1991,Lipowsky-PRL1994} the classical critical properties of the wetting layer and its thickness are severely affected by long-range dispersion forces.
Inclusion of  
dispersion interaction forces 
of the van der Waals type, or, equivalently, a proper treatment of the electromagnetic fluctuations 
implied by the different dielectric functions,  $\epsilon_1(\omega)$,  $\epsilon_2(\omega)$ and  $\epsilon_3(\omega)$ of the exterior of the system, of the metastable 
and of the stable phases respectively,~\cite{Dzyaloshinskii1961,Ginzburg1975} introduces between the surface ($z=0$) 
and the wetting interface ($z= l$) an effective additional interaction of the general form  $A/l^2$ where A is  the so-called Hamaker constant. 
When roughly speaking $\epsilon_1 <\epsilon_2 <\epsilon_3$ then $A>0$, the two interfaces
repel, and the wetting layer thickness divergence is greatly boosted, turning from logarithmic to power 
law~\cite{Lipowsky-PRB1985}
 \be
l \sim |h-h_c|^{-1/3},\label{l_power}
\ee
and, consequently, 
\be
\xi_{||} \sim |h-h_c|^{-2/3}.\label{xi_power}
\ee
That will be the case, for instance, for a slab in contact with the vacuum where the stable phase 
is a correlated metal and the metastable one a Mott insulator. 
Remarkably instead, in the alternative case $\epsilon_2 >\epsilon_3$ and $\epsilon_2 >\epsilon_1$, the growth of the wetting film is blocked, no longer diverging at the bulk transition.
That may occur for example when the stable bulk is Mott insulating, and the metastable state is metallic. 
Both circumstances are realized, e.g., in the classical self-wetting of crystals near the triple point, that is the physics of surface melting, where enhanced power-law growth of the liquid film, or alternatively its blocking and surface nonmelting, are well known to occur.~\cite{Tartaglino2005}.

As already mentioned, an anomalously thick {\sl dead layer} has been observed in photoemission at the surface of metallic V$_2$O$_3$ close to the first order transition to the paramagnetic insulator.~\cite{Marino2009} This phenomenon was explained in Ref.~\onlinecite{Borghi-PRL} studying the interface between the vacuum and a correlated metal described by the half-filled single-band Hubbard model at zero temperature. At $T=0$ the single-band Hubbard model displays a continuous Mott transition, which entails the existence of a correlation length $\xi_\text{bulk}$ that diverges at the transition. In the calculation it is just such a critical length that controls the thickness of the dead layer, which can therefore be very large, in accordance with experiments. However, the finite temperature Mott transition in V$_2$O$_3$ is first order so that, besides $\xi_\text{bulk}$, which does not diverge anymore, there will be another characteristic length, the thickness $l$ of the wetting layer, Eq.~\eqn{conclusion-xi_perp} or, better, Eq.~\eqn{l_power}, which instead may still show critical behavior approaching the transition. It would be interesting to further explore experimentally this issue to assess the actual origin and properties of the observed dead layer.     

We finally mention that another ingredient that should be taken into account but which we have disregarded is disorder. It is known that disorder can play a very important role in classical wetting.~\cite{Kardar-1985,Lipowsky-disorder,Fisher-1986-Faraday,Kardar&Hao-1990,FORGACS} However, in the presence of disorder one cannot export anymore the results of classical wetting in $d+1$ dimensions to the quantum case in $d$ dimensions, since the disorder is constant in the time direction hence has generically stronger effects. Even more, we expect that also in the absence of a surface the properties of the quantum model around the transition may be altered by impurities~\cite{TVojta-2014}. However, disorder is just one of the issues that need detailed investigation to get more accurate and realistic descriptions. Among the others, we mention in connection with first-order Mott transitions the fate of the interfacial modes Eq.~\eqn{conclusion-omega} and their observable tracks in the particular case where one of the two phases, specifically the metal, is gapless. 



\section*{Acknowledgments}
This work has been supported by 
the European Union, Seventh Framework Programme, under the project
GO FAST, grant agreement no. 280555, and under ERC MODPHYSFRICT grant no.320796.

 
 \appendix
 \section{Spin-wave approximation}
\label{appendix-1}
One can formally derive the Landau-Ginzburg Hamiltonian Eq.~\eqn{H-fluct-3d} also through the quantum Ising model Eq.~\eqn{Ham} treated by spin-wave approximation. We consider the spin model 
\bea
\mathcal{H} &=& - \sum_{i,j}\,\gamma_{i,j}\,\sigma^z_i\,\sigma^z_j 
-\sum_{ijkl} \tilde{\gamma}_{i,j;k,l}\,\sigma^z_i\,\sigma^z_j \,\sigma^z_k\,\sigma^z_l \nonumber\\
&& - h\sum_i\,\sigma^x_i
\equiv P[\sigma^z] - h\sum_i\,\sigma^x_i,\label{new-Ham}\\
\eea
where the site label $i=(z_i,\br_i)$ and $P[m]$ is a polynomial operator of the $\sigma^z_i$'s. We assume that 
$\gamma_{i,j}=\gamma_{j,i}$ 
and $
\tilde{\gamma}_{i,j;k,l}=\tilde{\gamma}_{j,i;k,l}=
\tilde{\gamma}_{k,l;i,j},
$ are translationally invariant,  
and  perform the unitary transformation 
\be
\mathcal{U} = \exp\Bigg[-\fract{i}{2}\,\sum_i\, \theta_i\,\sigma^y_{i}\Bigg],
\ee
with Euler angle $\theta_i=\theta(z_i)\in[0,\pi]$ appropriate to the slab geometry, so that 
\ba
\mathcal{U}^\dagger\,\sigma^z_{i}\,\mathcal{U} &=& 
\cos\theta_i\,\sigma^z_{i} - \sin\theta_i\,\sigma^x_{i},\\
\mathcal{U}^\dagger\,\sigma^x_{i}\,\mathcal{U} &=& 
\cos\theta_i\,\sigma^x_{i} + \sin\theta_i\,\sigma^z_{i}, 
\ea
and the Hamiltonian transforms into
\bea
\mathcal{U}^\dagger\,\mathcal{H}\,\mathcal{U} 
&=& P\Big[\big\{\cos\theta_i\,\sigma^z_i- 
\sin\theta_i\,\sigma^x_{i}\big\}\Big]
\nonumber\\
&& - h\sum_i\,\Big(\cos\theta_i\,\sigma^x_{i} + \sin\theta_i\,\sigma^z_{i}\Big).\label{app-transformed-Ham}
\eea
We shall assume that, after the unitary transformation, the ground state of Eq.~\eqn{app-transformed-Ham} is very close to the fully polarised state with 
$\sigma^z_i=1$, $\forall i$. We thus write
\ba
\sigma^z_i &\simeq& 1 - \big(x_i^2 + p_i^2 -1
\big) \equiv 1 - \Pi_i,\\
\sigma^x_i &\simeq& \sqrt{2}\,x_i,\\
\sigma^y_i &\simeq& \sqrt{2}\, p_i,
\ea
where $x_i$ and $p_i$ are conjugate variables, under the assumption that $\langle \Pi_i\rangle \ll 1$ and the vacuum of the bosonic operators $b^\dagga_i$ and $b^\dagger_i$ associated 
to $x_i$ and $p_i$ is the fully polarised state. This amounts to 
impose that the transformed Hamiltonian has no term linear in $x_i$ or $p_i$, a condition that actually fixes the Euler angles $\theta_i$ by the set of equations
\bea 
h\,\cos\theta_i &=& 
-\fract{\delta P[m]}{\delta m_i}_{\big|\{m_j\}=\{\cos\theta_j\}}\;\sin\theta_i,\label{app-conditions}
\eea   
which reads explicitly
\bea
h\,\cot\theta_i &=& 
2\sum_j\,\gamma_{ij}\,\cos\theta_j \label{eq-sw}\\
&&+ 4\sum_{jkl}\,\tilde{\gamma}_{i,j;k,l}\,\cos\theta_j\,\cos\theta_k\,\cos\theta_l,\nonumber
\eea
to be solved with appropriate boundary conditions. 
We observe that Eq.~\eqn{eq-sw} is the saddle point equation of  the classical energy functional 
\bea
E[m] &=& - \sum_{i,j}\,\gamma_{i,j}\,m_i\,m_j 
- h\sum_i\,\sqrt{1-m_i^2} \nonumber\\
&& -\sum_{ijkl} \tilde{\gamma}_{i,j;k,l}\,m_i\,m_j\,m_k\,m_l,\label{classical-E} 
\eea
under the substitution $m_i=\cos\theta_i$.
Next, we expand the transformed Hamiltonian 
$\mathcal{U}^\dagger\,\mathcal{H}\,\mathcal{U}\to \mathcal{H}$
at first order in $\Pi_i$ and second in $x_i$, the first order vanishing 
by Eq.~\eqn{app-conditions},
\bea
\mathcal{H} &=& P\Big[\big\{\cos\theta_i
-\cos\theta_i\,\Pi_i - 
\sqrt{2}\;\sin\theta_i\,x_{i}\big\}\Big]\nonumber
\\
&& - h\sum_i\,\Big(\sqrt{2}\;\cos\theta_i\,x_{i} 
+ \sin\theta_i -\sin\theta_i\,\Pi_{i}\Big)\nonumber\\
&=& E[m] +\sum_i\,
\left(h\,\sin\theta_i-\fract{\delta P[m]}{\delta m_i}\,\cos\theta_i
\right)\,\Pi_i \nonumber\\
&& + \sum_{ij}\,\fract{\delta^2 P[m]}{\delta m_i\delta m_j}\,
\sin\theta_i\,\sin\theta_j\, x_i\,x_j\nonumber\\
&=& E[m] +\sum_i\,\fract{h}{\sin\theta_i}\,
\Big(x_i^2 + p_i^2 -1\Big)\nonumber\\
&& + \sum_{ij}\,\fract{\delta^2 P[m]}{\delta m_i\delta m_j}\,
\sin\theta_i\,\sin\theta_j\, x_i\,x_j\nonumber\\
&=& E[m] - \sum_i\, \fract{h}{\sin\theta_i} 
+ \sum_i\, \fract{h}{\sin\theta_i}\;p_i^2  \nonumber\\  
&& + \sum_{i,j}\,\fract{\partial^2 E[m]}{\partial m_i\,\partial m_j}\;\sin\theta_i\,\sin\theta_j\,
x_i\,x_j.\label{H-sw} 
\eea
After the canonical transformation 
\ba
x_i &\to& \fract{1}{\sqrt{2}\;\sin\theta_i}\,x_i,\\
p_i &\to&  \sqrt{2}\;\sin\theta_i\;p_i,
\ea
Eq.~\eqn{H-sw} becomes 
\bea
\mathcal{H} &=& E[m] - \sum_i\, \fract{h}{\sin\theta_i} 
+ 2h\,\sum_i\, \sin\theta_i\;p_i^2  \nonumber\\  
&& + \frac{1}{2}\,\sum_{i,j}\,\fract{\partial^2 E[m]}{\partial m_i\,\partial m_j}\;x_i\,x_j,
\eea
which is what we would get by the Hamiltonian 
\[
\mathcal{H}_* = 2h\,\sum_i\, \sqrt{1-m_i^2}\,\sin^2 p_i + E[m],
\]
with $m_i$ and $p_i$ conjugate variables, expanded at second order in the deviation from the saddle point solution $p_{i\text{sp}}=0$ and $m_{i\text{sp}}=\cos\theta_i$.
In the continuum limit we thus obtain exactly the same equations we used in the main body of the text.
 
In order to formally diagonalize the Hamiltonian it is however more convenient to perform the transformation 
\ba
x_i &\to& \sqrt{\fract{2}{\sin\theta_i}}\;\,x_i,\\
p_i &\to& \sqrt{\fract{\sin\theta_i}{2}}\;p_i,
\ea 
so that 
\ba
\mathcal{H} &=& E[m] - \sum_i\, \fract{h}{\sin\theta_i} 
+ \fract{h}{2}\,\sum_i\, p_i^2  \\  
&& \!\!+ \frac{1}{2}\,\sum_{i,j}\,4\,\fract{\partial^2 E[m]}{\partial m_i\,\partial m_j}\;
\sqrt{\sin\theta_i}\;\sqrt{\sin\theta_j}\;
x_i\,x_j.
\ea
The quadratic form that appears in the last term can be diagonalized 
\bea
&&\sum_j\, 4\,\fract{\partial^2 E[m]}{\partial m_i\,\partial m_j}\;
\sqrt{\sin\theta_i}\;\sqrt{\sin\theta_j}\; \psi_{n,j} = 
\epsilon_n\,\psi_{n,i}\qquad \qquad ,\label{app-E2}
\eea
with appropriate boundary conditions imposed on the wavefunctions $\psi_{n,i}$, 
so that, after the unitary transformation 
\[
x_i = \sum_n\,\psi_{n,i}\,x_n,\qquad
p_i = \sum_n\,\psi_{n,i}\,p_n,
\]
the Hamiltonian reads
\ba
\mathcal{H} &=& E[m] - \sum_i\, \fract{h}{\sin\theta_i} \\  
&& \!\!+ \fract{h}{2}\,\sum_n\, p_n^2   + \frac{1}{2}\,\sum_{n}\,\epsilon_n\,
x_n^2,
\ea
that can be brought to the diagonal form
\be
\mathcal{H} = E[m] - \sum_i\, \fract{h}{\sin\theta_i} 
+ \sum_{n}\,\omega_n\,\left(b^\dagger_n b^\dagga_n + \frac{1}{2}\right)
\ee
by the transformation 
\[
x_n \to \sqrt{\fract{h}{\epsilon_n}}\;\,x_n,\qquad 
p_n \to \sqrt{\fract{\epsilon_n}{h}}\;p_n,
\]
with eigenvalues 
\be
\omega_n = \sqrt{h\,\epsilon_n}.
\ee
An advantage of this derivation is that it provides the quantum fluctuation correction to the classical energy. 
Indeed, if $\omega_n$ are the eigenvalues of the harmonic part, the ground state energy is
\be
E_0 = E[m] - \sum_i\, \fract{h}{\sin\theta_i} + \fract{1}{2}\sum_n\,\omega_n.
\ee

\bibliography{Nonequilibrium}
\bibliographystyle{apsrev}

\end{document}